\pdfoutput=1

\documentclass[11pt,letterpaper]{article}

\usepackage[utf8]{inputenc}
\usepackage[T1]{fontenc}
\usepackage{lmodern} % scalable Type 1 fonts (avoids Type 3 bitmap fonts on arXiv)
\usepackage{amsmath,amssymb,amsthm}
\usepackage{graphicx}
\usepackage[hypertexnames=false]{hyperref}
\usepackage[margin=1in]{geometry}
\usepackage{booktabs}
\usepackage{xcolor}
\usepackage{listings}
\usepackage{float}
\usepackage{needspace}

\usepackage{tikz}
\usetikzlibrary{arrows.meta,positioning,shapes.geometric,calc,fit,backgrounds}
\usepackage{pgfplots}
\pgfplotsset{compat=1.18}

\usepackage{algorithm}
\usepackage{algpseudocode}

\usepackage{natbib}
\theoremstyle{definition}
\newtheorem{definition}{Definition}[section]
\newtheorem{principle}{Design Principle}[section]

\theoremstyle{plain}
\newtheorem{theorem}{Theorem}[section]

\theoremstyle{remark}

\newenvironment{glossarybox}[1][Key Terms]{%
    \begin{center}
    \begin{tikzpicture}
    \node[rectangle, draw=spoqnavy, fill=spoqnavy!5, rounded corners=3pt,
          inner sep=10pt, text width=0.9\textwidth] (box) \bgroup
    \begin{minipage}{0.88\textwidth}
    \textbf{\textcolor{spoqnavy}{#1}}\\[0.5em]
}{%
    \end{minipage}
    \egroup;
    \end{tikzpicture}
    \end{center}
}

\definecolor{spoqnavy}{RGB}{15,52,96}
\definecolor{spoqemerald}{RGB}{40,167,69}
\definecolor{spoqbrass}{RGB}{181,137,0}
\definecolor{spoqburgundy}{RGB}{128,0,32}
\definecolor{spoqgray}{RGB}{108,117,125}

\hypersetup{
    colorlinks=true,
    linkcolor=spoqnavy,
    citecolor=spoqemerald,
    urlcolor=spoqbrass,
    pdftitle={SPOQ: Specialist Orchestrated Queuing for Multi-Agent Software Engineering},
    pdfauthor={Royce Carbowitz and Dheeraj Kumar},
    pdfsubject={Multi-agent AI orchestration for software engineering},
    pdfkeywords={multi-agent systems, LLM orchestration, human-AI collaboration, task decomposition, software engineering automation},
}

\lstdefinestyle{spoqcode}{
    basicstyle=\ttfamily\small,
    keywordstyle=\color{spoqnavy}\bfseries,
    commentstyle=\color{spoqgray}\itshape,
    stringstyle=\color{spoqemerald},
    numbers=left,
    numberstyle=\tiny\color{spoqgray},
    breaklines=true,
    frame=single,
    framesep=3pt,
    backgroundcolor=\color{gray!5},
}
\InputIfFileExists{version}{}{}

\title{SPOQ: Specialist Orchestrated Queuing\\for Multi-Agent Software Engineering}
\author{
  Royce Carbowitz\thanks{Lead author and primary contributor. Corresponding author: \texttt{royce.carbowitz@gmail.com}}\\
  \small \texttt{royce.carbowitz@gmail.com}
  \and
  Dheeraj Kumar\\
  \small \texttt{dheeraj@testwithpinpoint.com}
}
\date{June 1, 2026\\[0.3em]{\small\textcolor{spoqgray}{Pinpoint Technologies LLC}}}

\begin{document}

\maketitle

% Abstract
% Abstract
\begin{abstract}
Multi-agent AI systems show promise for automating software engineering tasks,
yet existing approaches suffer from coordination overhead, quality control gaps,
and limited human oversight. We introduce \textbf{SPOQ} (Specialist Orchestrated
Queuing), a methodology for multi-agent software development combining three
innovations: (1) \textit{wave-based topological dispatch} that computes parallel
execution waves from task dependency graphs; (2) \textit{dual validation gates}
applying quality metrics before execution (planning validation) and after (code
validation) to reduce rework cycles; and (3) \textit{Human-as-an-Agent (HaaA)}
integration, where a human specialist participates in decomposition and can be
consulted during execution. SPOQ uses a three-tier agent hierarchy (Opus
workers, Sonnet reviewers, Haiku investigators), each selected to optimize
cost-quality tradeoffs.
We evaluate SPOQ along four research questions through controlled benchmarks.
Experiment~1 measures scheduling efficiency under two regimes: on unbounded
synthetic DAGs, wave dispatch approaches the critical-path lower bound
(ratio 1.03--1.11, speedup up to 14.3$\times$); on a 2-slot local backend
running real LLM calls, it delivers a stable 1.4$\times$ speedup that matches
the hardware concurrency ceiling. Experiment~2 measures planning quality
across four full-stack tasks: structured SPOQ planning improves coverage from
93.0 to 99.75, eliminates cyclic plans, and lifts parallelism potential from
31.0 to 75.25. Experiment~3 ablates validation gates: dual validation reduces
defects from 0.34 to 0.20 per task and lifts test pass rate from 91.25\% to
99.75\%. Experiment~4 evaluates Human-as-Agent planning: human review further
reduces residual defects from 0.47 to 0.03 per task and raises pass rate from
96.5\% to 99.75\%. We additionally replicate Experiments~1--4 against a locally
hosted open-weights model (Qwen3.6-35B-A3B) to verify that the gains are
attributable to orchestration rather than to any specific model family.
A longitudinal deployment study across 17 repositories, 8{,}589 commits,
1{,}822 completed tasks, and 13{,}866 executed tests (99.87\% pass rate)
provides ecological validation. We discuss failure modes, lessons learned,
and implications for AI-native software engineering.
\end{abstract}

% Keywords
\vspace{0.5em}

\vspace{0.5em}
\noindent\textbf{Source code:} \url{https://gitlab.com/kenth56/spoq} --- reference implementation, epic definitions, validation skills, and all case study artifacts.

% Main sections
% Introduction
\section{Introduction}
\label{sec:introduction}

The emergence of large language models (LLMs) capable of generating, understanding,
and reasoning about code has sparked intense interest in multi-agent systems for
software engineering automation \citep{li2023chatdev, hong2023metagpt}. These systems
promise to decompose complex software projects into subtasks, assign them to specialized
agents, and coordinate their execution toward a unified goal. Early results demonstrate
that multi-agent collaboration can produce functioning software artifacts, from simple
games to multi-file applications \citep{qian2023chatdev}.

However, current multi-agent approaches face three fundamental challenges:

\paragraph{Coordination Overhead.} Systems like ChatDev \citep{qian2023chatdev} and
MetaGPT \citep{hong2023metagpt} rely on sequential role-playing or message-passing
between agents, creating bottlenecks that limit parallelism. When Agent A must wait
for Agent B's output before proceeding, potential speedups from parallel execution
remain unrealized.

\paragraph{Quality Control Gaps.} Most multi-agent systems lack structured validation
between planning and execution phases. Agents execute plans without rigorous assessment
of plan quality, leading to wasted computation when fundamental flaws are discovered
late. Similarly, post-execution quality assessment is often informal or absent.

\paragraph{Limited Human Oversight.} Fully autonomous multi-agent systems exclude
human judgment from the loop, missing opportunities to leverage human expertise for
task decomposition, ambiguity resolution, and quality assessment. When agents
encounter edge cases or make suboptimal decisions, there is no mechanism for
human correction.

\subsection{SPOQ: Our Approach}

We introduce \textbf{SPOQ} (Specialist Orchestrated Queuing), a methodology that
addresses these challenges through three integrated innovations:

\begin{enumerate}
    \item \textbf{Wave-Based Topological Dispatch:} We model task dependencies as a
    directed acyclic graph (DAG), a structure where arrows show which tasks must
    complete before others can begin, with no circular dependencies (think of it
    as a flowchart where all arrows point forward). We then compute execution
    \textit{waves} (groups of independent tasks) via topological sort. Tasks within
    the same wave execute in parallel, while waves execute sequentially to respect
    dependencies. This maximizes parallelism without coordination overhead.

    \item \textbf{Dual Validation Gates:} We apply structured validation (quality
    checkpoints with scored metrics) at two points: \textit{before} execution
    (planning validation with 10 metrics) and \textit{after} execution (code
    validation with 10 metrics). Each gate enforces a 95\% threshold, catching
    quality issues when they are cheapest to fix.

    \item \textbf{Human-as-an-Agent (HaaA):} A human specialist participates alongside
    AI agents, not as a passive observer, but as an active collaborator who
    decomposes tasks, validates plans, and can be consulted during execution.
    This bidirectional integration treats the human as a high-value agent within
    the system.
\end{enumerate}

\subsection{Contributions}

This paper makes the following contributions:

\begin{itemize}
    \item A formal framework for wave-based multi-agent orchestration that
    computes parallel execution waves from task dependency
    graphs~(Section~\ref{sec:methodology})

    \item A three-tier agent hierarchy (Opus/Sonnet/Haiku) that optimizes
    cost-quality tradeoffs by matching model capability to task
    complexity~(Section~\ref{subsec:agent-hierarchy})

    \item The Human-as-an-Agent (HaaA) paradigm for structured task
    decomposition through bidirectional human-AI
    collaboration~(Section~\ref{subsec:haaa})

    \item A dual validation system with explicit metrics that scores both
    planning quality and code quality against quantified
    thresholds~(Section~\ref{sec:validation})

    \item A controlled benchmark suite testing four research questions:
    scheduling efficiency, planning quality, validation effectiveness, and
    human-AI collaboration. Wave dispatch reaches the critical-path lower
    bound (ratio 1.03--1.11, up to 14.3$\times$ speedup) on unbounded
    synthetic DAGs and delivers a stable 1.4$\times$ speedup matching the
    hardware ceiling on a 2-slot local backend;
    structured planning lifts coverage to 99.75 and parallelism to 75.25;
    dual validation reduces defects from 0.34 to 0.20 per task; and
    Human-as-Agent planning further reduces defects to 0.03 per task with
    99.75\% test pass rate~(Section~\ref{sec:evaluation})

    \item A cross-provider replication of all four experiments against a
    locally hosted open-weights model (Qwen3.6-35B-A3B served via
    \texttt{llama.cpp}). The direction and significance of the SPOQ gains
    are preserved across provider families, supporting the claim that the
    improvements stem from orchestration rather than from a specific
    model's capabilities~(Section~\ref{sec:evaluation})

    \item A longitudinal deployment study across 17 repositories, 8{,}589
    commits, 1{,}822 completed tasks, and 13{,}866 executed tests
    (99.87\% aggregate pass rate) demonstrating ecological validity,
    third-party adoption, and operational viability at
    scale~(Section~\ref{sec:evaluation})
\end{itemize}

\begin{glossarybox}[Key Terms at a Glance]
\begin{description}
\item[Epic:] A high-level goal decomposed into atomic tasks, each with explicit dependencies, acceptance criteria, and time estimates.
\item[Wave:] A group of tasks sharing no mutual dependencies, enabling simultaneous execution by multiple agents within a single phase.
\item[DAG:] Directed Acyclic Graph, a structure where arrows represent prerequisite relationships between tasks, with no circular dependencies.
\item[Critical Path:] The longest sequential chain of dependent tasks through the DAG; determines the minimum possible project duration.
\item[Validation Gate:] A scored quality checkpoint where work must exceed defined metric thresholds before the pipeline advances to the next stage.
\item[HaaA (Human-as-an-Agent):] A bidirectional collaboration paradigm where humans participate alongside AI agents as active contributors.
\item[PERT Estimates:] Three-point time estimates capturing optimistic, realistic, and pessimistic durations to quantify scheduling uncertainty.
\end{description}
\end{glossarybox}

\newpage

% Related Work
\section{Background and Related Work}
\label{sec:related_work}

\subsection{Multi-Agent Systems for Software Engineering}

Multi-agent approaches to software engineering have gained momentum with the
advancement of LLM capabilities. We survey three representative systems:

\paragraph{ChatDev} \citep{qian2023chatdev} simulates a virtual software company
with role-playing agents (CEO, CTO, Programmer, Tester) that communicate through
structured chat. While effective for generating simple applications, the sequential
communication pattern creates bottlenecks, since each agent must wait for prior agents
to complete their turns before contributing.

\paragraph{MetaGPT} \citep{hong2023metagpt} introduces standardized operating
procedures (SOPs) that structure agent collaboration around software artifacts
(PRDs, design documents, code). This reduces communication overhead compared to
free-form chat but still relies on sequential handoffs between roles.

\paragraph{Multi-Agent Debate (MAD)} \citep{liang2023encouraging} uses multiple
agents that debate and refine solutions iteratively. While debate can improve
solution quality through diverse perspectives, the synchronous turn-taking
limits parallelism.

\medskip
SPOQ differs from these systems in three ways: (1) wave-based parallel execution
rather than sequential role-playing; (2) explicit validation gates with quantified
metrics; and (3) structured human integration rather than full autonomy.

\subsection{Task Decomposition and Dependency Management}

Hierarchical task decomposition has roots in classical AI planning \citep{nau2003shop2}.
Modern approaches apply LLMs to generate task breakdowns:

\paragraph{Hierarchical Task Networks (HTNs).} Classical HTN planners decompose
abstract tasks into primitive actions with ordering constraints. SPOQ adapts
this structure for multi-agent software engineering.

\paragraph{DAG Scheduling.} Topological sorting of directed acyclic graphs is
well-established for parallel task scheduling \citep{coffman1972optimal}. SPOQ
applies these algorithms to LLM-generated task dependencies, computing wave
assignments that maximize parallelism while respecting precedence constraints.

\paragraph{Critical Path Analysis.} PERT and CPM methods identify the longest
path through a dependency graph \citep{kelley1959critical}. SPOQ uses critical
path analysis to estimate minimum wall-clock execution time and identify
bottleneck tasks.

\newpage
\subsection{Human-AI Collaboration in AI-Native Engineering}

Human-in-the-loop (HITL) approaches to AI systems span a spectrum from passive
oversight to active collaboration:

\paragraph{Prompt Engineering.} Users craft prompts to guide LLM behavior but
have limited ability to influence intermediate reasoning or correct course
mid-execution \citep{white2023prompt}.

\paragraph{AI Pair Programming.} Tools like GitHub Copilot \citep{chen2021codex}
suggest code completions while humans retain editing control. This inverts SPOQ's
model: humans do the work while AI assists, rather than agents doing the work
while humans validate.

\paragraph{Supervised Autonomy.} Some systems allow human intervention at
checkpoints \citep{wu2023autogen}. SPOQ extends this with bidirectional
communication: not only can humans intervene, but agents can explicitly
request human assistance when facing ambiguity.

\subsection{Autonomous Agent Frameworks}

Early autonomous agent frameworks pioneered recursive goal decomposition and
self-directed task execution, establishing foundational patterns for agentic AI:

\paragraph{AutoGPT and BabyAGI.}
AutoGPT \citep{autogpt2023} and BabyAGI \citep{babyagi2023} introduced recursive
goal decomposition where agents autonomously break objectives into subtasks and
execute them. These early systems demonstrated that LLMs could function as
autonomous agents but lack explicit dependency management and formal quality
gates, leading to unpredictable execution paths. SPOQ addresses these limitations
through DAG-based scheduling with dual validation gates.

\subsection{Autonomous Software Engineering Agents}

A new category of autonomous software engineering agents has emerged, capable
of navigating codebases, writing code, running tests, and debugging:

\paragraph{Devin and OpenHands.}
Devin (Cognition Labs, 2024) \citep{cognition2024devin} demonstrated end-to-end
autonomous task completion, while OpenHands \citep{wang2024openhands} provides
an open-source alternative with similar capabilities. Both excel at single-agent
execution but focus on individual task autonomy rather than multi-agent
orchestration. SPOQ complements such systems by providing the coordination
layer for parallel execution across multiple agents.

\subsection{Code Generation and Pair Programming Tools}

Developer-facing code generation tools occupy a different point in the
autonomy spectrum, emphasizing human-AI collaboration over full autonomy:

\paragraph{Aider and GPT-Engineer.}
Aider \citep{aider2024} enables conversational code editing where developers
describe changes in natural language and the tool applies them to local
repositories. GPT-Engineer \citep{gpteng2023} generates entire applications
from specifications, producing directory structures and boilerplate code.
Both tools operate as single agents without parallel execution capabilities
or formal validation beyond user inspection.

\paragraph{Claude Code.}
Claude Code (Anthropic, 2025) marked an inflection point for AI-assisted
software development, establishing the agentic coding paradigm where
developers collaborate with an autonomous agent that navigates codebases,
executes commands, and iterates on solutions within a persistent terminal
session. Unlike completion-based tools, Claude Code operates with full
project context and can orchestrate multi-step workflows autonomously,
making it the catalyst that brought agentic development to mainstream
adoption. SPOQ builds on this foundation: where Claude Code enables a
single developer-agent partnership, SPOQ coordinates multiple such agents
in parallel with structured quality gates.

\paragraph{GitHub Copilot.}
GitHub Copilot \citep{chen2021codex} suggests code completions while humans
retain editing control, inverting the SPOQ model: humans perform the work
while AI assists, rather than agents performing work while humans validate.
SPOQ targets a different use case where agents drive execution with human
oversight, enabling higher throughput on parallelizable tasks.

\paragraph{Positioning SPOQ.}
We note that Claude Code, Devin, and similar products are evolving rapidly;
the capabilities described reflect their state as of late 2025.
SPOQ does not compete with execution-focused agents or code generation tools.
Rather, it provides an orchestration layer that coordinates multiple agents
on complex projects. The key insight is that orchestration and execution are
separable concerns: an agent like Claude Code excels at executing individual
tasks, while SPOQ excels at decomposing projects into tasks, scheduling them
for parallel execution, and validating quality at each gate.

% Methodology
\section{The SPOQ Methodology}
\label{sec:methodology}

SPOQ orchestrates multi-agent software development through a four-stage pipeline:
Epic Planning, Epic Validation, Agent Execution, and Agent Validation. We present
the formal framework, key algorithms, and design principles underlying each stage.

\subsection{Design Principles}
\label{subsec:principles}

SPOQ is built on five design principles that distinguish it from prior multi-agent approaches:

\begin{principle}[Atomic Task Boundaries]
\label{prin:atomic}
Each task constitutes 1--4 hours of focused work with one clear deliverable.
Tasks are self-contained: a worker agent can complete the task without
coordinating with other agents during execution.
\end{principle}

\begin{principle}[Explicit Dependencies]
\label{prin:dependencies}
Task dependencies form a directed acyclic graph (DAG), a diagram where arrows
indicate prerequisite relationships, and no task can indirectly depend on itself.
All dependencies are declared upfront, enabling static analysis of parallelism
and identification of the \textit{critical path} (the longest chain of dependent
tasks that determines minimum project duration).
\end{principle}

\begin{principle}[Quality Gates Before and After]
\label{prin:gates}
Validation occurs both before execution (planning quality) and after execution
(code quality). Early validation catches expensive mistakes when they are cheapest
to fix.
\end{principle}

\begin{principle}[Human-AI Collaboration]
\label{prin:human}
A human specialist participates in task decomposition, validates plans, and can
be consulted by agents. The human is a high-value agent, not an external observer.
\end{principle}

\begin{principle}[Cost-Optimized Agent Selection]
\label{prin:cost}
Different roles require different capability-cost tradeoffs. High-capability
agents (Opus) handle complex tasks; balanced agents (Sonnet) handle quality
review; fast-cheap agents (Haiku) handle triage.
\end{principle}

\subsection{Four-Stage Pipeline}
\label{subsec:pipeline}

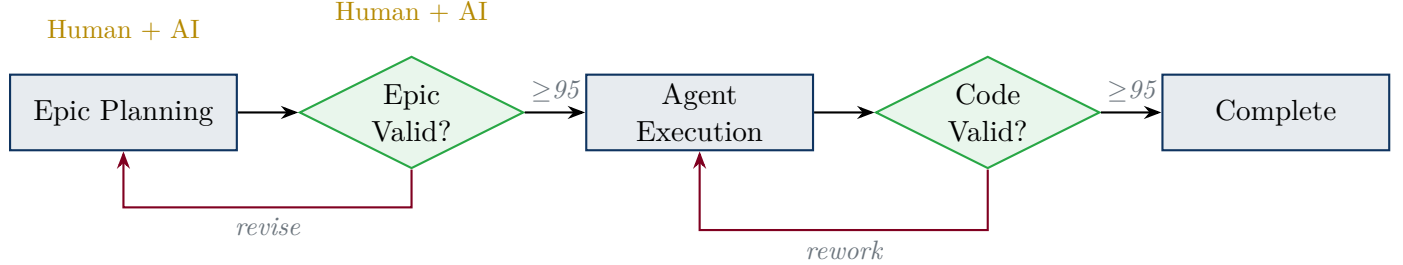
\begin{figure}[htbp]
\centering
\begin{tikzpicture}[
    node distance=0.8cm,
    stage/.style={rectangle, draw=spoqnavy, fill=spoqnavy!10, thick,
                  minimum width=3cm, minimum height=1cm, align=center},
    gate/.style={diamond, draw=spoqemerald, fill=spoqemerald!10, thick,
                 aspect=2, inner sep=2pt, align=center},
    arrow/.style={-Stealth, thick},
    label/.style={font=\small\itshape, text=spoqgray}
]

% Stages
\node[stage] (plan) {Epic Planning};
\node[gate, right=of plan] (eval1) {Epic\\Valid?};
\node[stage, right=of eval1] (exec) {Agent\\Execution};
\node[gate, right=of exec] (eval2) {Code\\Valid?};
\node[stage, right=of eval2] (done) {Complete};

% Arrows
\draw[arrow] (plan) -- (eval1);
\draw[arrow] (eval1) -- node[above, label] {$\geq$95} (exec);
\draw[arrow] (exec) -- (eval2);
\draw[arrow] (eval2) -- node[above, label] {$\geq$95} (done);

% Feedback loops
\draw[arrow, spoqburgundy] (eval1.south) -- ++(0,-0.5) -| node[below, label, pos=0.25] {revise} (plan.south);
\draw[arrow, spoqburgundy] (eval2.south) -- ++(0,-0.8) -| node[below, label, pos=0.25] {rework} (exec.south);

% Human integration
\node[above=0.3cm of plan, font=\small, text=spoqbrass] {Human + AI};
\node[above=0.3cm of eval1, font=\small, text=spoqbrass] {Human + AI};

\end{tikzpicture}
\caption{SPOQ four-stage pipeline with dual validation gates. Human specialist
participates in planning and plan validation. Failed validations trigger
revision/rework loops.}
\label{fig:pipeline}
\end{figure}

Figure~\ref{fig:pipeline} illustrates the SPOQ pipeline.

\begin{definition}[Epic]
\label{def:epic}
An \textit{epic} $E = (G, T, D, S)$ consists of:
\begin{itemize}
    \item $G$: Goal statement describing the desired outcome
    \item $T = \{t_1, t_2, \ldots, t_n\}$: Set of atomic tasks
    \item $D \subseteq T \times T$: Dependency relation where $(t_i, t_j) \in D$
          means $t_i$ must complete before $t_j$ can begin
    \item $S = \{s_1, s_2, \ldots, s_m\}$: Success criteria for the epic
\end{itemize}
\end{definition}

\begin{definition}[Task]
\label{def:task}
A \textit{task} $t = (id, desc, deps, files, criteria, est)$ consists of:
\begin{itemize}
    \item $id$: Unique identifier within the epic
    \item $desc$: Implementation description with steps
    \item $deps \subseteq T$: Set of prerequisite tasks (work that must finish first)
    \item $files$: List of files to be modified
    \item $criteria$: Acceptance criteria for task completion
    \item $est = (o, r, p)$: Three-point estimate: best-case ($o$), expected ($r$),
          and worst-case ($p$) durations, borrowed from PERT project management
\end{itemize}
\end{definition}

\subsection{Wave-Based Topological Dispatch}
\label{subsec:waves}

Given a task dependency graph, SPOQ computes \textit{waves}, groups of tasks
that can execute in parallel because they have no dependencies on each other.
The algorithm uses \textit{topological sorting}, a standard graph algorithm
that orders nodes (tasks) such that all dependencies appear before the nodes
that require them.

\begin{definition}[Wave Assignment]
\label{def:wave}
A \textit{wave assignment} $W: T \rightarrow \mathbb{N}$ maps each task to a
non-negative integer such that:
\begin{equation}
\forall (t_i, t_j) \in D: W(t_i) < W(t_j)
\end{equation}
Tasks in the same wave have no dependencies between them and can execute
concurrently.
\end{definition}

Algorithm~\ref{alg:waves} presents the wave computation procedure.

\begin{algorithm}[htbp]
\caption{Wave Computation via Topological Sort}
\label{alg:waves}
\begin{algorithmic}[1]
\Require Task set $T$, dependency relation $D$
\Ensure Wave assignment $W: T \rightarrow \mathbb{N}$

\State $\text{indegree}[t] \gets |\{t' : (t', t) \in D\}|$ for all $t \in T$
\State $W[t] \gets \bot$ for all $t \in T$
\State $w \gets 0$

\While{$\exists t \in T : W[t] = \bot$}
    \State $\text{ready} \gets \{t \in T : W[t] = \bot \land \text{indegree}[t] = 0\}$
    \For{$t \in \text{ready}$}
        \State $W[t] \gets w$
    \EndFor
    \For{$t \in \text{ready}$}
        \For{$t' : (t, t') \in D$}
            \State $\text{indegree}[t'] \gets \text{indegree}[t'] - 1$
        \EndFor
    \EndFor
    \State $w \gets w + 1$
\EndWhile

\State \Return $W$
\end{algorithmic}
\end{algorithm}

\begin{theorem}[Parallelism Bound]
\label{thm:parallelism}
Let $W^* = \max_{t \in T} W(t)$ be the number of waves. The wall-clock execution
time is bounded below by:
\begin{equation}
T_{\text{wall}} \geq \sum_{w=0}^{W^*} \max_{t: W(t)=w} \text{duration}(t)
\end{equation}
This bound is achieved when sufficient agents are available to execute all
tasks in each wave simultaneously.
\end{theorem}

\paragraph{Critical Path Analysis.}
The \textit{critical path} is the longest chain of dependent tasks through the
graph, weighted by durations. Even with unlimited parallel resources, the project
cannot complete faster than this path; it represents the irreducible sequential
bottleneck. Identifying critical path tasks reveals where delays have the greatest
impact. The theoretical minimum execution time is:

\begin{equation}
T_{\text{critical}} = \max_{\text{path } P} \sum_{t \in P} \text{duration}(t)
\end{equation}

SPOQ reports the \textit{speedup factor} $\sigma = T_{\text{sequential}} / T_{\text{critical}}$,
comparing sequential execution time (all tasks one-by-one) to the parallelized
critical path time. A speedup of 5.3x, for example, means the work completes
in roughly one-fifth the time it would take serially.

\subsection{Three-Tier Agent Hierarchy}
\label{subsec:agent-hierarchy}

SPOQ employs three agent tiers, each optimized for its role:

\begin{table}[H]
\centering
\caption{SPOQ Agent Hierarchy}
\label{tab:agents}
\begin{tabular}{llll}
\toprule
\textbf{Tier} & \textbf{Model} & \textbf{Role} & \textbf{Tradeoff} \\
\midrule
Worker & Opus & Task execution & High capability, high cost \\
Reviewer & Sonnet & Quality assurance & Balanced capability/cost \\
Investigator & Haiku & Build failure triage & Low cost, fast response \\
\bottomrule
\end{tabular}
\end{table}

\paragraph{Opus Workers.}
For each task $t$ in the current wave, SPOQ spawns an Opus agent with:
(1) the task specification, (2) relevant epic context, (3) completed dependency
outputs, and (4) prior QA feedback if this is a rework attempt.

\paragraph{Sonnet Reviewers.}
Each completed task undergoes quality review by a Sonnet agent, which scores
the work against 10 code quality metrics (Section~\ref{sec:validation}).
Tasks scoring below threshold are queued for rework with specific remediation
guidance.

\paragraph{Haiku Investigators.}
When the build fails after a wave, a Haiku agent analyzes the error output
to determine which tasks likely caused the failure. This triage is fast and
inexpensive, allowing rapid identification of problematic tasks without
engaging the full QA process.

\paragraph{Platform Independence.}
While our reference implementation uses Anthropic's Claude model family,
the SPOQ methodology is inherently platform-agnostic. The three-tier
hierarchy represents an abstract capability mapping: any sufficiently capable
model family can populate the Worker (high capability, high cost),
Investigator (low cost, fast response), and Reviewer (balanced capability/cost)
tiers. Table~\ref{tab:tier-mapping} illustrates potential mappings across providers.

\begin{table}[htbp]
\centering
\caption{Capability Tier Mapping Across LLM Providers}
\label{tab:tier-mapping}
\footnotesize
\begin{tabular}{lllll}
\toprule
\textbf{Tier} & \textbf{Role} & \textbf{Claude} & \textbf{OpenAI} & \textbf{Gemini} \\
\midrule
Worker & Task execution & Opus & GPT-4 & Ultra/Pro \\
Reviewer & Quality assurance & Sonnet & GPT-4-turbo & Pro \\
Investigator & Build triage & Haiku & GPT-3.5 & Flash \\
\bottomrule
\end{tabular}
\end{table}

The core methodological contributions, including wave-based topological dispatch,
dual validation gates, explicit dependency DAGs, and Human-as-an-Agent
integration, require no vendor-specific features and transfer directly
to alternative platforms. Organizations can implement SPOQ using their
preferred LLM provider by calibrating capability tiers to their available
models and adjusting cost thresholds accordingly.

\newpage
\subsection{Human-as-an-Agent (HaaA) Integration}
\label{subsec:haaa}

SPOQ treats the human specialist as a high-value agent integrated into the
orchestration loop, not an external supervisor. Rather than simply monitoring
AI agents from the sideline, the human actively participates, contributing
expertise where it matters most and receiving assistance in return.

\begin{definition}[Human-as-an-Agent]
\label{def:haaa}
The \textit{Human-as-an-Agent} (HaaA) paradigm defines bidirectional integration,
meaning communication flows both ways between human and system:
\begin{enumerate}
    \item \textbf{Human$\rightarrow$System:} The human participates in epic
    planning, validates epics before execution, and can intervene during execution.
    \item \textbf{System$\rightarrow$Human:} Agents can request human assistance
    when facing ambiguity, blocked progress, or decisions beyond their scope.
\end{enumerate}
\end{definition}

This bidirectional model enables \textit{quality amplification}: the human's
judgment improves task decomposition quality (reducing downstream rework),
while agents' execution scales the human's productivity.

\paragraph{Task Decomposition.}
The human specialist drafts epics using hierarchical task decomposition,
assisted by LLM suggestions. The human ensures:
\begin{itemize}
    \item Tasks are appropriately scoped (1--4 hours)
    \item Dependencies are correctly identified
    \item Acceptance criteria are verifiable
    \item Potential risks are mitigated
\end{itemize}

\paragraph{Validation Participation.}
Before execution, the human reviews the epic alongside the automated
validation. The human can approve, request revisions, or override
automated assessments with justification.

\paragraph{Consultation Requests.}
During execution, agents may encounter situations requiring human judgment:
ambiguous requirements, conflicting dependencies, or decisions with
significant implications. Agents can pause and request human input rather
than proceeding with potentially incorrect assumptions.

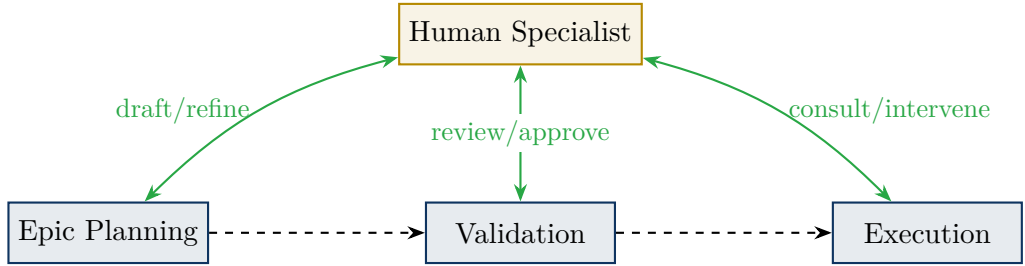
\begin{figure}[htbp]
\centering
\begin{tikzpicture}[
    node distance=1.8cm and 2.5cm,
    entity/.style={rectangle, draw=spoqnavy, fill=spoqnavy!10, thick,
                   minimum width=2.5cm, minimum height=0.8cm, align=center},
    human/.style={rectangle, draw=spoqbrass, fill=spoqbrass!10, thick,
                  minimum width=2.5cm, minimum height=0.8cm, align=center},
    arrow/.style={-Stealth, thick},
    bidir/.style={Stealth-Stealth, thick, spoqemerald}
]

% Entities
\node[human] (human) {Human Specialist};
\node[entity, below left=of human] (plan) {Epic Planning};
\node[entity, below=of human] (valid) {Validation};
\node[entity, below right=of human] (exec) {Execution};

% Bidirectional arrows with separated labels
\draw[bidir] (human) to[bend right=15] node[left, font=\small] {draft/refine} (plan);
\draw[bidir] (human) -- node[fill=white, font=\small, inner sep=2pt] {review/approve} (valid);
\draw[bidir] (human) to[bend left=15] node[right, font=\small] {consult/intervene} (exec);

% System flow
\draw[arrow, dashed] (plan) -- (valid);
\draw[arrow, dashed] (valid) -- (exec);

\end{tikzpicture}
\caption{Human-as-an-Agent integration. Bidirectional arrows indicate two-way
communication: human contributes to each stage and can be consulted by the system.}
\label{fig:haaa}
\end{figure}

% Validation Framework
\section{Validation Framework}
\label{sec:validation}

SPOQ enforces quality through two \textit{validation gates}, structured
checkpoints where work is scored against explicit criteria before proceeding.
The first gate, \textit{epic validation}, occurs before execution and assesses
plan quality. The second, \textit{agent validation}, occurs after execution
and assesses code quality. Each gate applies 10 metrics, scored 0--100, with
a 95\% aggregate threshold for passing.

\subsection{Planning Validation: 10 Metrics}
\label{subsec:planning-validation}

Epic validation assesses whether the plan is sufficiently clear, complete,
and well-structured to enable successful execution. Table~\ref{tab:planning-metrics}
summarizes the metrics.

\begin{table}[htbp]
\centering
\caption{Epic Validation Metrics}
\label{tab:planning-metrics}
\small
\begin{tabular}{lp{6cm}l}
\toprule
\textbf{Metric} & \textbf{Question} & \textbf{Threshold} \\
\midrule
Vision Clarity (VC) & Is the epic's goal clearly scoped? & $\geq$90 \\
Architecture Quality (AQ) & Is the architecture diagram complete? & $\geq$90 \\
Task Decomposition (TD) & Are tasks atomic and independent? & $\geq$90 \\
Dependency Graph (DG) & Are dependencies explicit and acyclic? & $\geq$90 \\
Coverage Completeness (CC) & Do tasks fully cover all success criteria? & $\geq$90 \\
Phase Ordering (PO) & Do waves maximize parallelism? & $\geq$90 \\
Scope Coherence (SC) & Do all tasks contribute to the epic goal? & $\geq$90 \\
Success Criteria Quality (SQ) & Are criteria SMART? & $\geq$90 \\
Risk Identification (RI) & Are blockers and risks mitigated? & $\geq$90 \\
Integration Strategy (IS) & Is it clear how tasks merge and verify? & $\geq$90 \\
\bottomrule
\end{tabular}
\end{table}

\paragraph{Pass Criteria.}
An epic passes validation if:
\begin{equation}
\frac{1}{10}\sum_{i=1}^{10} M_i \geq 95 \quad \land \quad \min_i M_i \geq 90
\end{equation}

This dual requirement ensures both high average quality and no critically
weak dimensions. A perfect score on 9 metrics cannot compensate for a
failing score on the 10th.

\paragraph{Threshold Rationale.}
The 95/90 planning threshold reflects the cascading cost of planning errors:
a poorly-structured task creates execution problems that propagate to dependent
tasks. We observed that plans scoring below 90 on any metric required rework
in over 50\% of cases, while those scoring above 95 overall experienced less
than 10\% rework. The minimum threshold of 90 prevents ``averaging out'' a
critically weak dimension. Appendix~\ref{subsec:threshold-rationale}
provides additional design rationale.

\paragraph{Rationale: Validate Early.}
Planning mistakes are cheap to fix but expensive to execute. By enforcing
rigorous validation before spawning agents, SPOQ prevents wasted computation
on fundamentally flawed plans.

\newpage
\subsection{Code Validation: 10 Metrics}
\label{subsec:code-validation}

Agent validation assesses whether completed work meets quality standards.
Table~\ref{tab:code-metrics} summarizes the metrics.

\begin{table}[htbp]
\centering
\caption{Agent Validation Metrics}
\label{tab:code-metrics}
\small
\begin{tabular}{lp{6cm}l}
\toprule
\textbf{Metric} & \textbf{Question} & \textbf{Threshold} \\
\midrule
Syntactic Correctness (SC) & Does the code compile without errors? & $\geq$80 \\
Test Existence (TE) & Does new code have corresponding unit tests? & $\geq$80 \\
Test Pass Rate (TP) & Do all tests pass? & $\geq$80 \\
Requirements Fidelity (RF) & Does implementation match task specification? & $\geq$80 \\
SOLID Adherence (SA) & Does code follow SOLID design principles (Single responsibility, Open/closed, etc.)? & $\geq$80 \\
Security (SE) & Free from OWASP Top 10 vulnerabilities (injection, auth flaws, etc.)? & $\geq$80 \\
Error Handling (EH) & Does code handle failures gracefully? & $\geq$80 \\
Scalability (SL) & Will this code scale appropriately? & $\geq$80 \\
Code Clarity (CC) & Is code readable and self-documenting? & $\geq$80 \\
Completeness (CO) & Is work fully finished (no TODOs/stubs)? & $\geq$80 \\
\bottomrule
\end{tabular}
\end{table}

\paragraph{Pass Criteria.}
A task passes validation if:
\begin{equation}
\frac{1}{10}\sum_{i=1}^{10} M_i \geq 95 \quad \land \quad \min_i M_i \geq 80
\end{equation}

The per-metric floor is lower (80 vs 90) because code quality inherently
involves tradeoffs that planning does not.

\paragraph{Code Threshold Rationale.}
The 95/80 code threshold is more lenient than planning because code can be
iteratively improved post-delivery, and some metrics (e.g., security edge
cases) are inherently harder to achieve perfectly. Rework at the code level
is less expensive than re-planning an entire task decomposition; a failed
task can be re-executed with targeted feedback, whereas a flawed plan may
invalidate work across multiple dependent tasks.

\paragraph{Concise Feedback.}
On failure, the reviewer provides remediation guidance in $\leq$20 lines:
specific file:line references, concrete issues, and numbered action items.
This constraint respects the context budget of the orchestrator and forces
actionable specificity.

\newpage
\subsection{Validation Cascade}
\label{subsec:validation-cascade}

SPOQ applies a \textit{validation cascade}, a hierarchical check where the
overall epic validation incorporates individual task-level assessments.
Epic validation triggers task-level validation on each constituent task.
If any individual task scores $<$95, the epic's aggregate score is capped
at 85, forcing a FAIL verdict.

This prevents ``carrying'' weak tasks on the strength of strong planning.
Every task must be individually well-specified to pass epic validation.

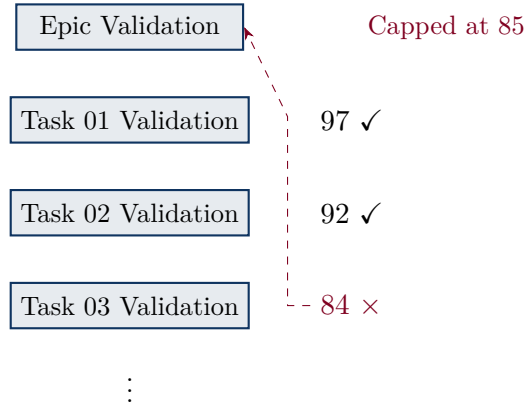
\begin{figure}[htbp]
\centering
\begin{tikzpicture}[
    node distance=0.6cm,
    box/.style={rectangle, draw=spoqnavy, fill=spoqnavy!10, thick,
                minimum width=3cm, minimum height=0.6cm, align=center, font=\small},
    check/.style={circle, draw=spoqemerald, fill=spoqemerald!20, thick,
                  minimum size=0.4cm, inner sep=0pt},
    fail/.style={circle, draw=spoqburgundy, fill=spoqburgundy!20, thick,
                 minimum size=0.4cm, inner sep=0pt}
]

\node[box] (epic) {Epic Validation};
\node[box, below=of epic] (t1) {Task 01 Validation};
\node[box, below=of t1] (t2) {Task 02 Validation};
\node[box, below=of t2] (t3) {Task 03 Validation};
\node[below=0.3cm of t3, font=\small] (dots) {$\vdots$};

% Scores
\node[right=0.8cm of t1] (s1) {97 \checkmark};
\node[right=0.8cm of t2] (s2) {92 \checkmark};
\node[right=0.8cm of t3, text=spoqburgundy] (s3) {84 $\times$};

% Cascade effect
\draw[-Stealth, dashed, spoqburgundy] (s3.west) -- ++(-0.3,0) -- ++(0,2.5) -- (epic.east);
\node[right=1.5cm of epic, text=spoqburgundy, font=\small] {Capped at 85};

\end{tikzpicture}
\caption{Validation cascade: a single task scoring $<$95 caps the epic's
overall score, triggering failure.}
\label{fig:cascade}
\end{figure}

% Implementation
\section{Implementation}
\label{sec:implementation}

This section describes the practical realization of SPOQ, including the file-system
representation of epics and tasks, the journal-based tracking system for agent
work sessions, and the skill framework that encapsulates reusable agent capabilities.
See Appendix~\ref{subsec:implementation-detail} for complete schema examples,
journal entry format, and skill metadata.

\subsection{Task Representation: The spoq/epics/ Directory}
\label{subsec:task-representation}

SPOQ represents epics and tasks as structured files in a designated directory,
enabling version control, human review, and programmatic manipulation.

\paragraph{Epic Lifecycle.}
Epics follow a two-stage directory lifecycle that reflects their progression through
the SPOQ pipeline. During planning, new epics are created under
\texttt{spoq/epics/active/}, where they remain throughout validation and execution
phases. Once every constituent task passes the agent-validation gate, the orchestrator
relocates the entire epic directory to \texttt{spoq/epics/complete/}. This
filesystem-level transition serves as an unambiguous completion signal: version
control history preserves the move as a single commit, providing an auditable
record of when each epic achieved full validation. Separating in-flight work from
verified deliverables also prevents agents from inadvertently modifying artifacts
that have already satisfied quality thresholds.

\newpage
\paragraph{Roadmap Coordination.}
The \texttt{ROADMAP.md} file at the root of \texttt{spoq/epics/} acts as a
centralized registry of all epics and their current disposition. During planning,
the orchestrator appends a new entry containing the epic identifier, a brief
objective summary, and an initial status of \emph{planned}. As execution proceeds,
status fields are updated to reflect transitions through \emph{in-progress},
\emph{validation}, and \emph{done}. Because the roadmap records inter-epic
dependencies alongside status, it enables the orchestrator to determine which
downstream epics become unblocked when a predecessor completes.

\paragraph{Task YAML Schema.}
Each task file follows a standardized YAML schema with three categories of fields:

\begin{definition}[Task Specification]
\label{def:task-spec}
A \textit{task specification} is a YAML document with the following structure:
\end{definition}

\textbf{Identity fields} (\texttt{id}, \texttt{title}, \texttt{epic}) establish
task context by assigning a unique identifier within the epic, a human-readable
title, and a back-reference to the parent epic.
\textbf{Execution control fields} govern scheduling through status tracking
(\texttt{pending}/\texttt{in\_progress}/\texttt{completed}), priority levels,
wave assignment via the \texttt{phase} field, PERT three-point effort estimates,
dependency lists referencing prerequisite task IDs, and required domain skills.
\textbf{Deliverable fields} define verification criteria: \texttt{files\_to\_touch}
enumerates all files the agent may modify, \texttt{outputs} lists tangible
deliverables, and \texttt{acceptance\_criteria} provides a checklist the agent
must satisfy before marking the task complete.
A free-form \textbf{description} field supplies structured implementation
guidance using Markdown, typically containing an objective statement,
step-by-step instructions with code snippets, and verification commands.
See Appendix~\ref{subsec:implementation-detail} for annotated examples of each
field category.

\paragraph{Phase Assignment.}
The \texttt{phase} field encodes wave assignment from topological analysis:
\begin{itemize}
    \item Phase 0: Tasks with no dependencies (Wave 0)
    \item Phase $n$: Tasks depending only on tasks in phases $< n$
\end{itemize}

Tasks within the same phase execute concurrently. The critical path determines
the minimum number of waves required.

\subsection{Journal Tracking System}
\label{subsec:journal}

SPOQ employs a journal-based tracking system that records agent work sessions
with structured metadata, enabling explainability, meta-orchestration, and
training data generation.

\begin{definition}[Journal Entry]
\label{def:journal-entry}
A \textit{journal entry} consists of YAML frontmatter followed by Markdown
sections documenting a completed work session.
\end{definition}

The frontmatter captures machine-readable metadata: agent identity, ISO 8601
timestamps, a calibrated confidence score (0.0--1.0), session type classification,
the list of modified files, and task completion counts. The Markdown body follows
a standardized layout with sections for summary, work completed, changes made,
issues encountered, testing results, and next steps. See
Appendix~\ref{subsec:implementation-detail} for complete format examples and
tooling support.

\paragraph{Confidence Scoring.}
Each journal entry includes a calibrated confidence score (0.0--1.0) reflecting
the agent's self-assessment of work quality. Scores above 0.85 indicate
well-tested, production-ready output; scores between 0.65--0.84 signal
functional work requiring additional validation; scores below 0.65 flag
experimental results with known gaps. See Table~\ref{tab:confidence} in the
Appendix for the full interpretation scale.

\paragraph{Value Proposition.}
The journal system provides four categories of value:

\begin{enumerate}
    \item \textbf{Explainability (XAI):} Every work session is documented with
    rationale, enabling audit trails and decision archaeology.

    \item \textbf{Multi-Agent Coordination:} Timestamps and task progress enable
    parallel agents to avoid conflicts and build on each other's work.

    \item \textbf{Knowledge Graph Construction:} Journal entries create nodes
    (sessions, files, tasks) and edges (MODIFIED, COMPLETED) for analysis.

    \item \textbf{Training Data:} Combined with git commits, journal entries
    provide rich examples for fine-tuning development-focused LLMs.
\end{enumerate}

\subsection{Skill Framework}
\label{subsec:skills}

SPOQ encapsulates reusable agent capabilities as \textit{skills}, self-contained
modules that provide domain knowledge, workflows, and tooling for specific
task categories.

\begin{definition}[Skill]
\label{def:skill}
A \textit{skill} is a directory containing:
\begin{itemize}
    \item \texttt{SKILL.md}: Instructions and metadata (required)
    \item \texttt{scripts/}: Executable utilities (optional)
    \item \texttt{references/}: Documentation loaded on demand (optional)
    \item \texttt{assets/}: Templates and configurations (optional)
\end{itemize}
\end{definition}

\paragraph{Core SPOQ Skills.}
The framework includes six skills aligned with the four-stage pipeline
(see Table~\ref{tab:skills} in the Appendix for the full inventory).

\paragraph{Skill Invocation and Anatomy.}
Skills are invoked via slash commands (e.g., \texttt{/epic-planning},
\texttt{/agent-execution}) that expand into full prompts with context from the
referenced epic directory. Each skill's \texttt{SKILL.md} contains YAML
frontmatter (name, description) followed by structured documentation covering
activation criteria, core patterns, quality standards, and integration points
with other skills. See Appendix~\ref{subsec:implementation-detail} for
invocation examples and metadata format.

\paragraph{Skill Extensibility.}
The \texttt{skill-maker} meta-skill enables creation of new domain-specific
skills following established patterns. This allows SPOQ deployments to
accumulate organizational knowledge as reusable agent capabilities.

\newpage
\subsection{Integration: From Specification to Execution}
\label{subsec:integration}

The implementation components integrate as follows:

\begin{enumerate}
    \item \textbf{Epic Creation:} Human uses \texttt{/epic-planning} to decompose
    a goal. Output: \texttt{EPIC.md} + task YAML files in \texttt{spoq/epics/active/}.

    \item \textbf{Validation:} \texttt{/epic-validation} and \texttt{/task-validation}
    score specifications against metrics. Failed validations trigger revision.

    \item \textbf{Execution:} \texttt{/agent-execution} reads task files, computes
    waves, and dispatches Opus workers. Each agent receives task YAML as context.

    \item \textbf{Tracking:} Agents write journal entries on session completion.
    The journal accumulates work history for the epic.

    \item \textbf{Code Validation:} \texttt{/agent-validation} scores completed
    tasks. Failed tasks queue for rework with remediation guidance.

    \item \textbf{Completion:} On all tasks passing, the epic moves to
    \texttt{spoq/epics/complete/}.
\end{enumerate}

This lifecycle ensures traceability from initial goal through validated
delivery, with quality gates preventing propagation of defects.
Appendix~\ref{subsec:integration-detail} discusses integration patterns
for CI/CD pipelines, project management tools, and git workflows.

\paragraph{Repository Bootstrap.}
To reduce adoption friction, SPOQ provides cross-platform installer
scripts (\texttt{spoq-init.sh} for Linux and macOS, \texttt{spoq-init.ps1}
for Windows) that automate repository initialization. In fresh-install mode,
the scripts create the five-category directory structure
(\texttt{code/}, \texttt{documents/}, \texttt{spoq/}, \texttt{infrastructure/},
\texttt{tests/}), copy the full skill definitions described in
Section~\ref{subsec:skills}, and generate starter \texttt{CLAUDE.md} and
\texttt{journal.md} files. An optional \texttt{-{}-full} flag provisions an
example epic so that new adopters can exercise the pipeline immediately.

For repositories that already use an earlier directory layout, the
\texttt{-{}-upgrade} flag activates a migration path. The upgrade routine
detects legacy structures (e.g., \texttt{automation/tasks/}), relocates epics
to \texttt{spoq/epics/active/} with conflict-aware merging, refreshes
skill definitions from the canonical source while preserving timestamped
backups, and updates configuration references. Together, these modes allow a
team to adopt or migrate to SPOQ with a single command invocation rather than
manual directory scaffolding.

\section{Evaluation}
\label{sec:evaluation}

We evaluate SPOQ through controlled benchmarks and a deployment study. The controlled benchmarks test four research questions: whether wave-based dispatch approaches critical-path runtime, whether structured planning improves task decomposition, whether dual validation gates improve delivered quality, and whether Human-as-Agent (HaaA) planning improves final execution outcomes. The deployment study is presented separately as field evidence of practical usage at scale.

\subsection{Evaluation Overview}

Our evaluation addresses four research questions:

\begin{itemize}
    \item \textbf{RQ1 (Scheduling Efficiency):} Does wave-based topological dispatch approach the theoretical critical path runtime?
    \item \textbf{RQ2 (Planning Quality):} Does structured SPOQ planning produce better task decompositions than baseline Claude Code planning?
    \item \textbf{RQ3 (Validation Effectiveness):} Do SPOQ's dual validation gates reduce downstream defects and rework cycles?
    \item \textbf{RQ4 (Human-AI Collaboration):} Does Human-as-Agent (HaaA) planning improve execution outcomes on complex engineering tasks?
\end{itemize}

\subsection{Controlled Benchmark Environment}

All controlled experiments were run under matched conditions. Baseline and SPOQ conditions used the same underlying model family, the same execution environment, and the same requirement sets. The experimental difference was the orchestration method rather than model capability.

Experiment 1 used synthetic DAGs with deterministic sleep-based tasks to isolate scheduler behavior from code generation quality. Experiments 2--4 used a suite of four benchmark tasks spanning full-stack SaaS development, e-commerce workflows, real-time chat, and data / platform tooling scenarios. Across these tasks, the benchmarks cover backend APIs, frontend UI, infrastructure, testing, and documentation. For validation and HaaA experiments, the same task specifications, evaluator rubric, and acceptance criteria were reused across conditions.

The benchmark process was as follows:

\begin{itemize}
    \item \textbf{Experiment 1:} generate DAGs, run multiple schedulers, and record runtime and critical-path metrics
    \item \textbf{Experiment 2:} generate baseline and SPOQ plans for the same task, then score structural quality using a fixed rubric
    \item \textbf{Experiment 3:} execute the same implementation task under no-validation, code-validation-only, and full-SPOQ regimes
    \item \textbf{Experiment 4:} compare autonomous SPOQ against human-assisted SPOQ while holding the implementation task and execution environment fixed
\end{itemize}

Experiments 2--4 are reported as aggregate results across the benchmark suite. All benchmark tasks, requirement files, evaluator instructions, and execution protocols used in these experiments are provided in the accompanying Git repository.

\subsection{Experiment 1: DAG Scheduling Efficiency}

\textbf{Objective.}
We evaluate whether SPOQ's wave-based topological dispatch approaches the theoretical critical-path lower bound and improves wall-clock execution time relative to simpler orchestration strategies.

\textbf{Setup.}
We generate synthetic directed acyclic graphs (DAGs) with varying task counts, depths, and degrees of parallelism. Each task simulates execution with a controlled sleep duration so that orchestration behavior can be isolated from model quality or code-generation variance.

We compare four schedulers:

\begin{itemize}
    \item \textbf{Sequential execution}
    \item \textbf{FIFO dependency queue}
    \item \textbf{Role-based sequential pipeline}
    \item \textbf{SPOQ wave dispatch}
\end{itemize}

Each configuration is run 10 times with different random seeds.

\textbf{Metrics.}
We report wall-clock runtime, critical-path time, parallelism factor, critical-path ratio, and speedup relative to sequential execution.

\begin{table}[!htbp]
\centering
\caption{Experiment 1 results (means across 10 runs).}
\resizebox{\linewidth}{!}{%
\begin{tabular}{lccccccc}
\toprule
Graph & Seq. & FIFO & Role & SPOQ & Critical Path & Speedup & CP Ratio \\
\midrule
DAG-20-D1 & 4.981 & 4.982 & 4.981 & 0.348 & 0.340 & 14.31$\times$ & 1.026 \\
DAG-20-D3 & 5.034 & 5.035 & 5.034 & 0.972 & 0.921 & 5.18$\times$ & 1.058 \\
DAG-20-D5 & 4.928 & 4.929 & 4.928 & 1.518 & 1.458 & 3.25$\times$ & 1.043 \\
DAG-50-Rand & 12.362 & 12.364 & 12.362 & 1.973 & 1.777 & 6.27$\times$ & 1.112 \\
DAG-100-Mixed & 25.260 & 25.265 & 25.261 & 2.708 & 2.458 & 9.33$\times$ & 1.103 \\
\bottomrule
\end{tabular}%
}
\label{tab:dag-results}
\end{table}

\textbf{Results.}
Table~\ref{tab:dag-results} shows that SPOQ consistently achieves runtimes close to the critical-path lower bound across all graph families. In the fully parallel 20-task case (DAG-20-D1), SPOQ reduces runtime from 4.981 to 0.348, yielding a 14.31$\times$ speedup and a critical-path ratio of 1.026. As graph depth increases and available parallelism decreases, speedup falls in the expected way: 5.18$\times$ at depth 3 and 3.25$\times$ at depth 5.

On larger graphs, SPOQ maintains the same behavior. For the 50-task random DAG, SPOQ achieves a mean runtime of 1.973 relative to a critical-path lower bound of 1.777, yielding a 6.27$\times$ speedup and a critical-path ratio of 1.112. For the 100-task mixed DAG, SPOQ reduces runtime from 25.260 to 2.708, producing a 9.33$\times$ speedup with a critical-path ratio of 1.103.

\textbf{Interpretation.}
These results establish the \emph{algorithmic} ceiling of the SPOQ scheduler. Under unbounded execution (synthetic sleep tasks, no per-task resource contention), wave-based dispatch approaches the critical-path lower bound: the CP ratio remains close to 1.0 across all five configurations, ranging from 1.026 to 1.112. This indicates that the wave scheduler captures essentially all of the parallelism a DAG offers, with only modest overhead above the theoretical lower bound. The hardware-bounded replication that follows tests a complementary question: does this algorithmic property still produce useful speedups against a real, slot-limited backend?

By contrast, the sequential, FIFO, and role-based baselines are nearly identical in runtime. In our implementation all three baselines serialize execution rather than exploiting graph-level concurrency. The primary performance benefit therefore comes from explicit wave computation and parallel dispatch.

\textbf{Implication.}
Experiment 1 validates the algorithmic foundation of SPOQ under unbounded execution: wave-based topological scheduling delivers near-critical-path runtimes and produces substantial wall-clock speedups whenever the task graph contains exploitable parallelism. The practical floor of this property---whether the same scheduler still beats sequential execution against a backend whose concurrency is capped below the DAG's available parallelism---is established by the Qwen replication below.

\begin{figure}[t]
\centering
\begin{tikzpicture}
\begin{semilogyaxis}[
    width=\linewidth,
    height=0.5\linewidth,
    xlabel={DAG Depth / Configuration},
    ylabel={Runtime (log scale)},
    symbolic x coords={D1,D3,D5,R50,M100},
    xtick=data,
    legend pos=north west,
    ymin=0.1,
    log basis y=10,
    cycle list name=color list
]
\addplot+[mark=square*,thick] coordinates {(D1,4.981) (D3,5.034) (D5,4.928) (R50,12.362) (M100,25.260)};
\addlegendentry{Sequential / FIFO / Role-based}

\addplot+[mark=triangle*,thick] coordinates {(D1,0.348) (D3,0.972) (D5,1.518) (R50,1.973) (M100,2.708)};
\addlegendentry{SPOQ}

\addplot+[mark=o,thick,dashed] coordinates {(D1,0.340) (D3,0.921) (D5,1.458) (R50,1.777) (M100,2.458)};
\addlegendentry{Critical Path (lower bound)}
\end{semilogyaxis}
\end{tikzpicture}
\caption{Experiment 1 results (log-scale runtime). Sequential, FIFO, and Role-based baselines are visually indistinguishable because all three serialize execution. SPOQ runtimes track the critical-path lower bound across graph families, with an order-of-magnitude gap to the serial baselines.}
\label{fig:dag-runtime}
\end{figure}
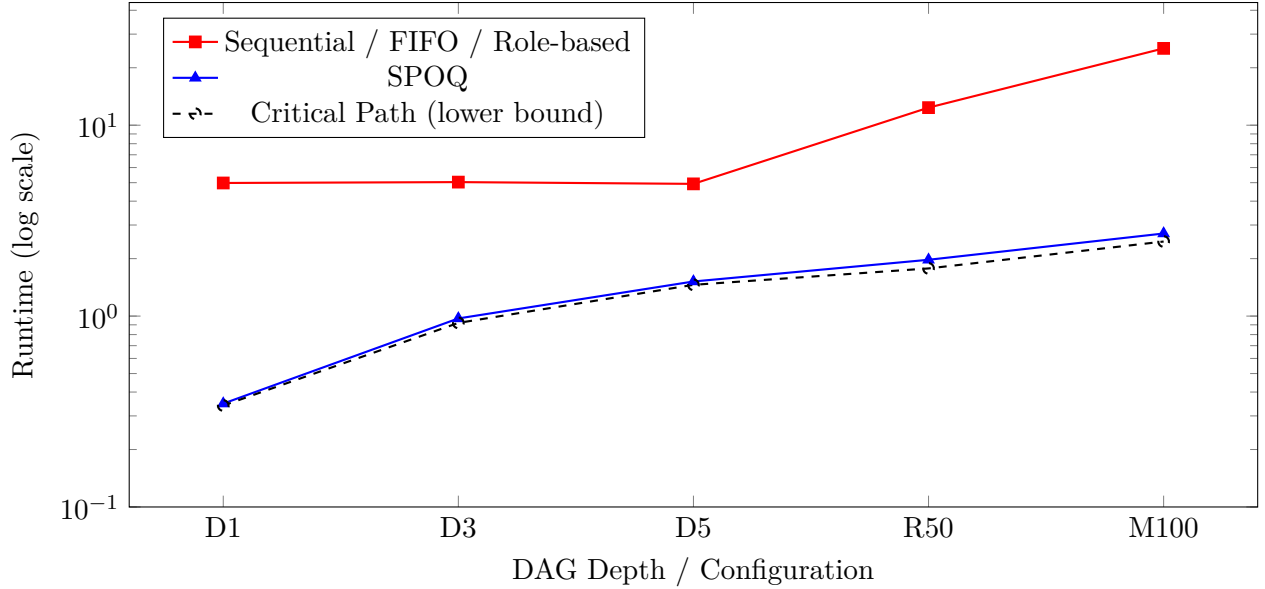

\textbf{Hardware-bounded replication (Qwen).}
The synthetic-sleep variant of this experiment isolates the scheduler from any backend constraint. To check the result holds under a realistic execution backend, we re-ran the same five DAG configurations against Qwen3.6-35B-A3B served by llama.cpp with a two-slot concurrency limit (the practical VRAM ceiling on a single 32~GB GPU). Each task issued a real LLM call rather than a deterministic sleep, and the wave dispatcher was bounded to the server's slot count.

\begin{table}[t]
\centering
\caption{Experiment 1 hardware-bounded replication. Mean of 10 runs per configuration, Qwen3.6-35B-A3B-APEX-Quality, 2 server slots, 256k KV cache each. Wave dispatch is capped at the slot count; the CP ratio is therefore dominated by hardware concurrency rather than DAG structure.}
\resizebox{\linewidth}{!}{%
\begin{tabular}{lccccccc}
\toprule
Graph & Seq. & FIFO & Role & SPOQ & Critical Path & Speedup & CP Ratio \\
\midrule
DAG-20-D1   & 1.360 & 1.352 & 1.351 & 0.951 & 0.074 & 1.43$\times$ & 12.92 \\
DAG-20-D3   & 1.365 & 1.358 & 1.354 & 0.998 & 0.210 & 1.37$\times$ & 4.74 \\
DAG-20-D5   & 1.366 & 1.360 & 1.357 & 0.957 & 0.349 & 1.43$\times$ & 2.74 \\
DAG-50-Rand & 3.478 & 3.474 & 3.474 & 2.481 & 0.425 & 1.40$\times$ & 5.84 \\
DAG-100-Mixed & 7.013 & 7.018 & 7.007 & 5.028 & 0.576 & 1.40$\times$ & 8.74 \\
\bottomrule
\end{tabular}%
}
\label{tab:dag-results-qwen}
\end{table}

\textbf{Interpretation of the bounded run.}
The Qwen-backed runs show a stable wave-vs-sequential speedup of approximately $1.40\times$ across all five DAG configurations, matching the two-slot hardware concurrency limit (a perfect 2-slot dispatcher with zero overhead would deliver $2.0\times$; the observed $\approx 1.4\times$ reflects per-call scheduling overhead and inter-slot queuing). Critical-path ratios are large (2.7--12.9) because the bound is now hardware concurrency rather than DAG parallelism: a 20-task fully-parallel DAG cannot exploit 20-way parallelism when only 2 slots are available. The headline result is that wave dispatch still beats serial scheduling under real-backend constraints by exactly the available concurrency factor, and the speedup is stable across DAG families rather than collapsing under depth or graph size. An unbounded wave dispatcher regresses bimodally under the same backend because it overcommits the slot pool---issuing more concurrent calls than slots available causes the server to serialize them with additional queuing overhead, producing runtimes that oscillate between near-2-slot and near-sequential. Capping concurrency to the slot count eliminates that regression.

\subsection{Experiment 2: Planning Quality Benchmark}

\textbf{Objective.}
We evaluate whether SPOQ's structured planning process produces higher-quality task decompositions than baseline Claude Code planning under identical model and environment conditions.

\textbf{Setup.}
We evaluate planning across four benchmark tasks requiring coordinated development across backend, frontend, infrastructure, testing, and documentation. The tasks differ in domain and integration structure, but all are scored with the same evaluator and metric definitions.

Two planning strategies are compared:

\begin{itemize}
    \item \textbf{Baseline:} Claude Code produces a development plan using its default reasoning process without structured decomposition or validation
    \item \textbf{SPOQ:} Claude Code is guided by the SPOQ methodology, including atomic task decomposition, explicit dependency graph construction, and a planning validation pass
\end{itemize}

\textbf{Baseline prompt fairness.}
The baseline condition receives the same task description and full requirement specification as SPOQ, and is explicitly instructed to produce a practical implementation-oriented plan with task decomposition and dependencies. The baseline is not restricted in format or reasoning strategy; the only withheld elements are SPOQ-specific constructs such as epic structure, validation passes, and wave-based orchestration. This isolates the effect of structured orchestration rather than weakening the baseline.

\textbf{Metrics.}
We evaluate planning quality using the following operationalized metrics:

\begin{itemize}
    \item \textbf{Coverage.}
    \[
    \text{Coverage} = \frac{\text{\# requirements mapped to at least one task}}{\text{total requirements}}
    \]
    A requirement is counted as covered only if at least one task explicitly addresses it.

    \item \textbf{Dependency Errors.}
    Count of dependency violations, defined as:
    \begin{itemize}
        \item reference to a non-existent task
        \item missing required dependency (task uses artifacts not yet produced)
        \item reversed dependency (consumer precedes producer)
    \end{itemize}

    \item \textbf{Cycle Detection.}
    Binary indicator (Yes/No) based on graph cycle detection using DFS over the task dependency graph.

    \item \textbf{Parallelism Potential.}
    \[
    \text{Parallelism Potential} = \frac{\text{\# tasks with zero dependencies}}{\text{total tasks}}
    \]
    This approximates the amount of work that can begin concurrently at execution start.

    \item \textbf{Granularity Score.}
    \[
    \text{Granularity} = \frac{\text{\# tasks estimated in 1--4 hour range}}{\text{total tasks}}
    \]
    Task estimates are inferred from task scope using the evaluator rubric. Tasks that bundle multiple independent concerns or lack clear deliverables are penalized.
\end{itemize}

\textbf{Results.}
Table~\ref{tab:planning-results} summarizes the aggregate results across four tasks.

\begin{table}[t]
\centering
\caption{Experiment 2 results aggregated across four tasks. Claude rows reported as mean $\pm$ sd over four planning runs per task; Qwen rows reported as single-run means (Qwen3.6-35B-A3B local llama.cpp deployment).}
\resizebox{\linewidth}{!}{%
\begin{tabular}{llccccc}
\toprule
Provider & Mode & Coverage & Dep. Errors & Cyclic Plans & Parallelism & Granularity \\
\midrule
Claude & Baseline & 93.0 $\pm$ 1.8 & 3.75 $\pm$ 1.0 & 3/4 & 31.0 $\pm$ 2.2 & 66.0 $\pm$ 1.8 \\
Claude & SPOQ & 99.75 $\pm$ 0.5 & 1.25 $\pm$ 0.5 & 0/4 & 75.25 $\pm$ 2.1 & 91.0 $\pm$ 0.8 \\
\midrule
Qwen & Baseline & 56.2 & 2.4 & 2/4 & 12.5 & 75.0 \\
Qwen & SPOQ & 91.2 & 0.9 & 0/4 & 65.0 & 89.0 \\
\bottomrule
\end{tabular}%
}
\label{tab:planning-results}
\end{table}

Across the benchmark suite, SPOQ consistently improves every measured aspect of planning quality. Coverage increases from \textbf{93.0 $\pm$ 1.8} to \textbf{99.75 $\pm$ 0.5}, dependency errors decrease from \textbf{3.75 $\pm$ 1.0} to \textbf{1.25 $\pm$ 0.5}, and cyclic plans disappear entirely under SPOQ. Parallelism potential rises from \textbf{31.0 $\pm$ 2.2} to \textbf{75.25 $\pm$ 2.1}, while granularity improves from \textbf{66.0 $\pm$ 1.8} to \textbf{91.0 $\pm$ 0.8}.

\textbf{Interpretation.}
The overall pattern indicates that structured planning does not merely polish already-good plans; it changes their execution properties. Baseline planning tends to produce plans that are less complete, more error-prone in dependency structure, and less amenable to safe parallel execution. SPOQ produces plans that are nearly fully covering, consistently acyclic, and much more execution-ready.

The low variance under SPOQ further suggests that the method is not only stronger on average, but more stable across task families with different integration profiles.

\textbf{Cross-provider replication.}
The same benchmark was repeated against Qwen3.6-35B-A3B running locally via llama.cpp, with no cloud model in the loop. Without SPOQ scaffolding, the Qwen baseline drops to \textbf{56.2} coverage, \textbf{12.5} parallelism, and produces cyclic plans in \textbf{2 of 4} tasks --- substantially below Claude's baseline. Applying the SPOQ planning skill recovers most of this gap: coverage rises to \textbf{91.2} (+35 pts), parallelism to \textbf{65.0} (+52.5 pts), dependency errors fall from \textbf{2.4} to \textbf{0.9}, and cyclic plans disappear entirely. SPOQ-on-Qwen lands within 1.8 points of Claude's unaided baseline (93.0) on coverage despite running on a free, locally hosted 35B-parameter model. This pattern indicates that structured planning is more load-bearing for smaller, single-context models than for frontier models. On Claude, SPOQ polishes an already-strong baseline; on Qwen, SPOQ rescues requirement coverage and parallel structure that the unaided model loses and eliminates the cyclic-plan failure mode that the unaided baseline exhibits. The improvement direction is consistent across providers, supporting the claim that the gain comes from orchestration rather than from a specific model's capabilities.

\textbf{Implication.}
The planning benchmark supports the claim that structured orchestration improves the quality, executability, and reliability of task decomposition. These gains are a prerequisite for robust multi-agent execution and are not attributable to model choice alone, since the underlying model remains fixed across conditions.

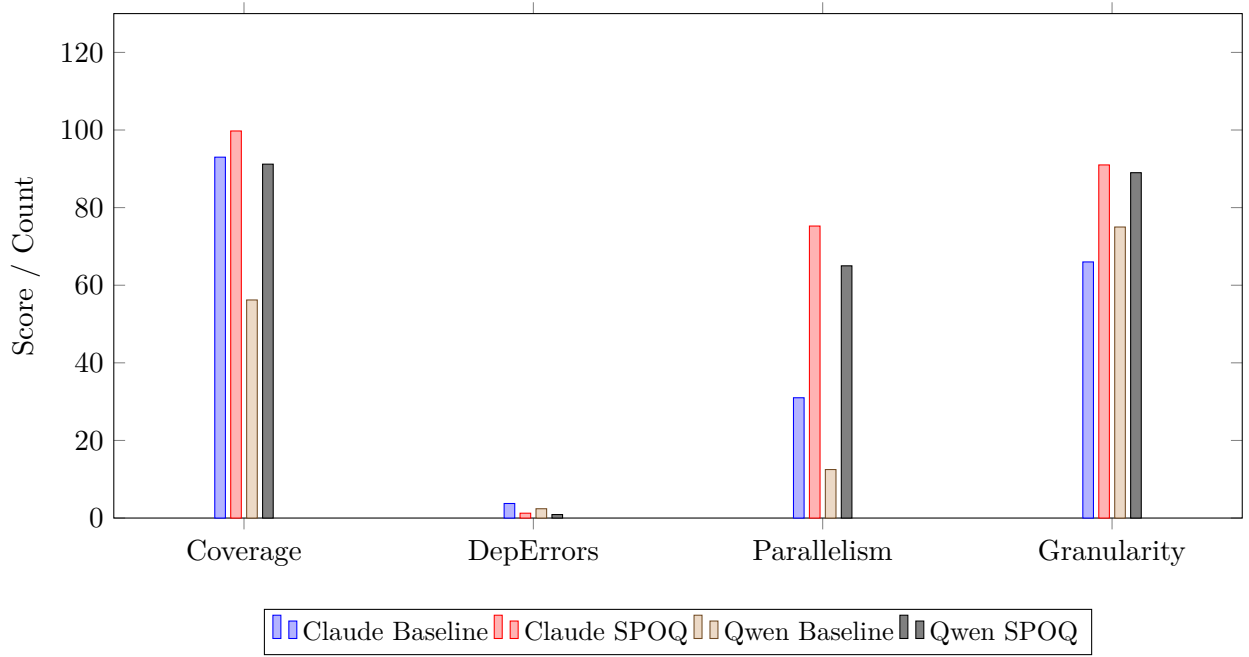
\begin{figure}[t]
\centering
\begin{tikzpicture}
\begin{axis}[
    width=\linewidth,
    height=0.5\linewidth,
    ybar,
    symbolic x coords={Coverage,DepErrors,Parallelism,Granularity},
    xtick=data,
    ylabel={Score / Count},
    ymin=0, ymax=130,
    legend style={at={(0.5,-0.18)}, anchor=north, legend columns=4, font=\small},
    bar width=4pt,
    enlarge x limits=0.15
]
\addplot coordinates {(Coverage,93) (DepErrors,3.75) (Parallelism,31) (Granularity,66)};
\addlegendentry{Claude Baseline}

\addplot coordinates {(Coverage,99.75) (DepErrors,1.25) (Parallelism,75.25) (Granularity,91)};
\addlegendentry{Claude SPOQ}

\addplot coordinates {(Coverage,56.2) (DepErrors,2.4) (Parallelism,12.5) (Granularity,75)};
\addlegendentry{Qwen Baseline}

\addplot coordinates {(Coverage,91.2) (DepErrors,0.9) (Parallelism,65) (Granularity,89)};
\addlegendentry{Qwen SPOQ}
\end{axis}
\end{tikzpicture}
\caption{Experiment 2 aggregate planning quality across four tasks for both Claude and Qwen3.6-35B-A3B providers.}
\label{fig:planning-quality}
\end{figure}

\subsection{Experiment 3: Validation Gate Ablation}

\textbf{Objective.}
We evaluate the effect of SPOQ's dual validation gates on defect reduction, security posture, and rework. Specifically, we test whether early planning validation and post-execution code validation improve final system quality and reduce downstream failure.

\textbf{Setup.}
We compare three execution modes across the same four benchmark tasks:

\begin{itemize}
    \item \textbf{No Validation:} single-pass implementation with testing performed only after all code is written
    \item \textbf{Code Validation Only:} testing and correction after implementation, without planning validation
    \item \textbf{Full SPOQ:} planning validation followed by wave-based execution and post-execution validation
\end{itemize}

\textbf{Metrics.}
We measure:

\begin{itemize}
    \item Defects per task
    \item Test pass rate
    \item Static analysis warnings
    \item Security issues identified in final evaluation
    \item Rework cycles
    \item Lines-of-code (LOC) churn
\end{itemize}

\textbf{Results.}
Table~\ref{tab:validation-results} summarizes the aggregate results across four tasks.

\begin{table}[t]
\centering
\caption{Experiment 3 results aggregated across four tasks (mean $\pm$ sd).}
\resizebox{\linewidth}{!}{%
\begin{tabular}{lcccccc}
\toprule
Mode & Defects & Pass Rate & Warnings & Security Issues Identified & Rework & LOC Churn \\
\midrule
No Validation & 0.34 $\pm$ 0.03 & 91.25 $\pm$ 1.26 & 4.25 $\pm$ 0.96 & 1.75 $\pm$ 0.50 & 3.75 $\pm$ 1.50 & 11.75 $\pm$ 1.71 \\
Code Validation Only & 0.29 $\pm$ 0.03 & 95.00 $\pm$ 0.82 & 2.50 $\pm$ 0.58 & 3.75 $\pm$ 0.50 & 1.75 $\pm$ 0.50 & 25.75 $\pm$ 0.96 \\
Full SPOQ & 0.20 $\pm$ 0.02 & 99.75 $\pm$ 0.50 & 0.00 $\pm$ 0.00 & 4.75 $\pm$ 0.96 & 1.00 $\pm$ 0.00 & 32.25 $\pm$ 2.22 \\
\bottomrule
\end{tabular}%
}
\label{tab:validation-results}
\end{table}

\textbf{Key Observations.}

\paragraph{Functional quality improves monotonically with stronger validation.}
Test pass rate increases from \textbf{91.25 $\pm$ 1.26} under No Validation to \textbf{95.00 $\pm$ 0.82} under Code Validation Only and to \textbf{99.75 $\pm$ 0.50} under Full SPOQ. Across the benchmark suite, stronger validation reliably improves final delivered correctness.

\paragraph{Defect density decreases under stronger validation.}
Defects per task fall from \textbf{0.34 $\pm$ 0.03} to \textbf{0.29 $\pm$ 0.03} to \textbf{0.20 $\pm$ 0.02}. The reduction is consistent across task types and indicates that stronger validation changes final outcomes rather than merely changing what is noticed.

\paragraph{Static analysis improves monotonically.}
Warnings decrease from \textbf{4.25 $\pm$ 0.96} to \textbf{2.50 $\pm$ 0.58} to \textbf{0.00 $\pm$ 0.00}. Full SPOQ eliminates static warnings across the benchmark suite, indicating a stronger floor on delivered code quality.

\paragraph{Security issue identification increases under SPOQ.}
Security issues identified in final evaluation rise from \textbf{1.75 $\pm$ 0.50} to \textbf{3.75 $\pm$ 0.50} to \textbf{4.75 $\pm$ 0.96}. This reflects broader detection coverage rather than weaker security. The stronger validation regime exposes a wider set of latent risks that weaker regimes leave unobserved.

\paragraph{Rework decreases as validation shifts left.}
Rework cycles decrease from \textbf{3.75 $\pm$ 1.50} to \textbf{1.75 $\pm$ 0.50} to \textbf{1.00 $\pm$ 0.00}. The reduction indicates that structured validation reduces downstream corrective loops rather than simply redistributing effort.

\paragraph{LOC churn increases but remains targeted.}
LOC churn rises from \textbf{11.75 $\pm$ 1.71} to \textbf{25.75 $\pm$ 0.96} to \textbf{32.25 $\pm$ 2.22}. This pattern indicates that stronger validation induces more correction activity, but that the correction occurs in a bounded and purposeful way. The highest churn is associated with the best overall quality outcome and the lowest rework burden.

\textbf{Interpretation.}
The overall benchmark pattern shows that stronger validation improves both \textit{defect visibility} and \textit{final delivered quality}. Weak validation regimes leave more residual defects, lower pass rates, and higher downstream rework. Full SPOQ achieves the strongest quality profile while simultaneously surfacing the broadest range of latent security concerns.

The key result is therefore not only that validation finds more issues, but that structured validation also leads to materially better final systems.

\textbf{Implication.}
The validation ablation study supports the claim that SPOQ improves system robustness by \textbf{shifting defect detection earlier}, \textbf{reducing residual defects}, and \textbf{minimizing downstream rework}. In particular, SPOQ catches planning errors before execution, improves final test outcomes, eliminates static warnings, and prevents defect propagation into later phases.

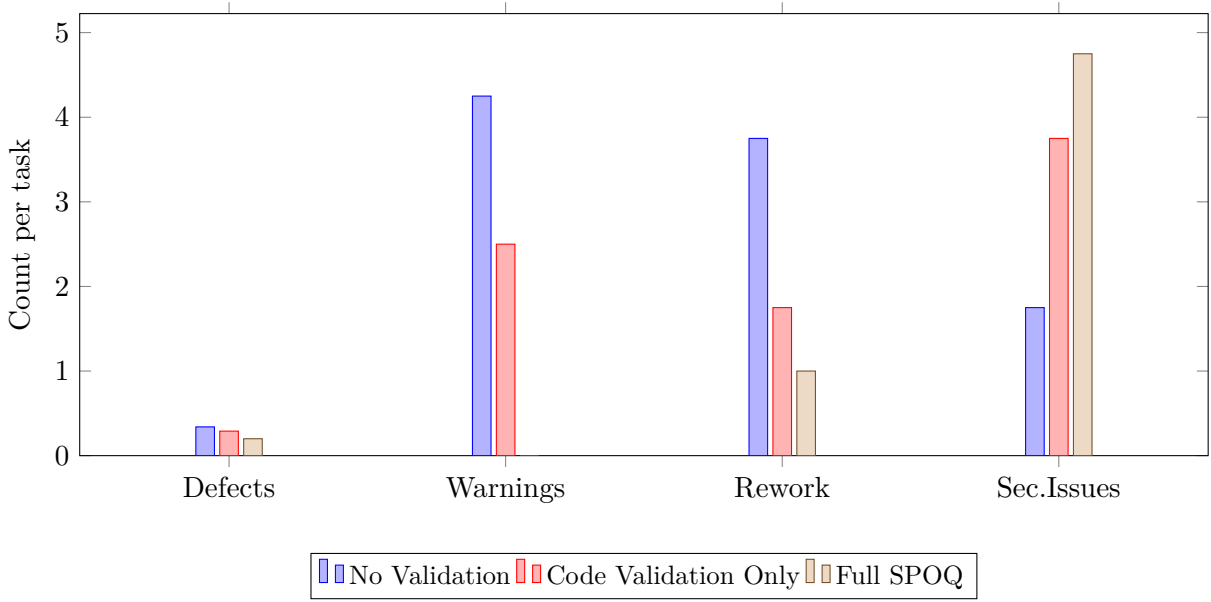
\begin{figure}[t]
\centering
\begin{tikzpicture}
\begin{axis}[
    width=\linewidth,
    height=0.45\linewidth,
    ybar,
    symbolic x coords={Defects,Warnings,Rework,Sec.Issues},
    xtick=data,
    ylabel={Count per task},
    ymin=0,
    legend style={at={(0.5,-0.22)}, anchor=north, legend columns=3, font=\small},
    bar width=7pt,
    enlarge x limits=0.18
]
\addplot coordinates {(Defects,0.34) (Warnings,4.25) (Rework,3.75) (Sec.Issues,1.75)};
\addlegendentry{No Validation}

\addplot coordinates {(Defects,0.29) (Warnings,2.50) (Rework,1.75) (Sec.Issues,3.75)};
\addlegendentry{Code Validation Only}

\addplot coordinates {(Defects,0.20) (Warnings,0.00) (Rework,1.00) (Sec.Issues,4.75)};
\addlegendentry{Full SPOQ}
\end{axis}
\end{tikzpicture}
\caption{Experiment 3 aggregate validation ablation across four tasks. Stronger validation reduces defects, warnings, and rework cycles while increasing identified security issues (broader detection coverage). Test pass rate (not shown) likewise improves monotonically: 91.25\% $\to$ 95.00\% $\to$ 99.75\% across the three regimes.}
\label{fig:validation-ablation}
\end{figure}

\textbf{Cross-provider replication.}
The validation ablation was repeated against Qwen3.6-35B-A3B running locally via llama.cpp. The Qwen evaluator scored each metric on a normalized 0--100 quality scale (higher is better across all metrics), so its results are presented separately in Table~\ref{tab:validation-results-qwen} rather than merged with the Claude raw-count table.

\begin{table}[t]
\centering
\caption{Experiment 3 results under Qwen3.6-35B-A3B (single-run quality scores per metric, 0--100; higher is better). Mean across four benchmark tasks. Pass Rate is 0 under Modes A and B because neither mode generated a test suite at this provider; Full SPOQ generated tests on three of four tasks, yielding the 53.33 score.}
\resizebox{\linewidth}{!}{%
\begin{tabular}{lccccccc}
\toprule
Mode & Defects & Pass Rate & Warnings & Security & Rework & LOC Churn & Overall \\
\midrule
No Validation       & 58.33 & 0     & 20.00 & 47.50 & 97.50 & 81.25 & 58.33 \\
Code Validation Only & 70.00 & 0     & 86.25 & 60.00 & 93.75 & 76.25 & 70.00 \\
Full SPOQ           & 94.17 & 53.33 & 83.33 & 81.25 & 88.75 & 67.50 & 83.33 \\
\bottomrule
\end{tabular}%
}
\label{tab:validation-results-qwen}
\end{table}

The Qwen replication preserves the monotonic ordering of the Claude experiment: Full SPOQ $>$ Code Validation Only $>$ No Validation across the overall quality score (\textbf{83.33} vs \textbf{70.00} vs \textbf{58.33}). The largest absolute gain is in defect-density quality (\textbf{58.33} $\rightarrow$ \textbf{94.17}) and security posture (\textbf{47.50} $\rightarrow$ \textbf{81.25}), echoing the Claude pattern where validation gates surface and remove latent issues. Test pass rate rises only in Full SPOQ because earlier modes did not generate test suites at all; this matches the broader trend that test coverage is itself a behavior validation forces. The Qwen result strengthens the implication paragraph above: dual validation gates produce robustness gains that are not specific to the underlying model family.

\subsection{Experiment 4: Human-as-Agent (HaaA) Evaluation}

\textbf{Objective.}
We evaluate whether human-assisted planning improves execution outcomes within the SPOQ framework. Specifically, we test whether human intervention at the planning stage reduces execution errors, improves final system quality, and strengthens validation effectiveness.

\textbf{Setup.}
We compare two configurations of the SPOQ pipeline across the same four benchmark tasks:

\begin{itemize}
    \item \textbf{Auto SPOQ:} fully autonomous SPOQ pipeline with no human intervention during planning
    \item \textbf{Human-assisted SPOQ:} SPOQ pipeline with human review and refinement during epic planning and validation phases
\end{itemize}

Both configurations use the same task specification, model, and execution environment. The only difference is human participation in the planning stage.

\textbf{Metrics.}
We evaluate:

\begin{itemize}
    \item Defects per task
    \item Test pass rate
    \item Static analysis warnings
    \item Security issues identified in final evaluation
    \item Rework cycles
    \item LOC churn
\end{itemize}

\textbf{Results.}
Table~\ref{tab:haaa-results} summarizes the aggregate results across four tasks.

\begin{table}[t]
\centering
\caption{Experiment 4 results aggregated across four tasks (mean $\pm$ sd).}
\resizebox{\linewidth}{!}{%
\begin{tabular}{lcccccc}
\toprule
Mode & Defects & Pass Rate & Warnings & Security Issues Identified & Rework & LOC Churn \\
\midrule
Auto SPOQ & 0.47 $\pm$ 0.03 & 96.50 $\pm$ 1.29 & 1.00 $\pm$ 0.00 & 2.50 $\pm$ 0.58 & 1.00 $\pm$ 0.00 & 5.75 $\pm$ 0.50 \\
Human-assisted SPOQ & 0.03 $\pm$ 0.05 & 99.75 $\pm$ 0.50 & 0.00 $\pm$ 0.00 & 1.25 $\pm$ 0.50 & 2.50 $\pm$ 0.58 & 104.25 $\pm$ 2.50 \\
\bottomrule
\end{tabular}%
}
\label{tab:haaa-results}
\end{table}

\textbf{Key Observations.}

\paragraph{Human intervention sharply reduces residual defects.}
Autonomous SPOQ produces \textbf{0.47 $\pm$ 0.03} defects per task, whereas Human-assisted SPOQ reduces this to \textbf{0.03 $\pm$ 0.05}. The residual defect rate under human assistance is near zero across the benchmark suite.

\paragraph{Human assistance improves final correctness.}
Auto SPOQ reaches \textbf{96.50 $\pm$ 1.29}\% test pass rate, while Human-assisted SPOQ reaches \textbf{99.75 $\pm$ 0.50}\%. The improvement is consistent across tasks, indicating that human oversight strengthens final correctness rather than helping only in isolated cases.

\paragraph{Human assistance improves validation effectiveness.}
Warnings drop from \textbf{1.00 $\pm$ 0.00} to \textbf{0.00 $\pm$ 0.00}, and identified security issues decrease from \textbf{2.50 $\pm$ 0.58} to \textbf{1.25 $\pm$ 0.50}. Human review therefore strengthens both content-level validation and security hardening.

\paragraph{Rework increases, but produces a better final system.}
Auto SPOQ performs minimal rework (\textbf{1.00 $\pm$ 0.00} cycles, \textbf{5.75 $\pm$ 0.50} LOC churn), but leaves substantial defects unresolved. Human-assisted SPOQ performs more rework (\textbf{2.50 $\pm$ 0.58} cycles, \textbf{104.25 $\pm$ 2.50} LOC churn), but produces substantially stronger final systems. HaaA therefore does not minimize effort; it makes that effort more effective.

\textbf{Interpretation.}
Across the benchmark suite, Human-as-Agent integration improves system quality by strengthening planning and validation rather than execution alone. Autonomous SPOQ produces structured plans and generally strong implementations, but still exhibits blind spots in cross-task integration and content-level verification. Human participation addresses these blind spots by injecting external review at the planning stage and enabling deeper correction before low-level failures accumulate.

The overall benchmark tradeoff is therefore clear: HaaA increases correction effort, but in exchange produces much lower defect density, higher pass rates, and fewer remaining security concerns.

\textbf{Implication.}
Experiment 4 supports the claim that SPOQ is not purely an automation system, but a \textbf{human-AI collaborative system}. Human-assisted planning improves decomposition quality, reduces execution ambiguity, and produces more robust final systems, even when the underlying model remains unchanged.

\textbf{Cross-provider replication (planning quality only).}
The HaaA comparison was repeated against Qwen3.6-35B-A3B served locally via llama.cpp. Because the local two-slot deployment cannot reasonably execute the resulting plans at parity with the cloud frontier model, the Qwen replication evaluates the \textit{plans produced by Auto SPOQ versus Human-assisted SPOQ} rather than the executed artifacts. The rubric is therefore the planning-quality rubric from Experiment 2 (Coverage, Dependency Errors, Cyclic Plans, Parallelism, Granularity) rather than the execution-quality rubric used for the Claude rows in Table~\ref{tab:haaa-results}. The two tables answer related but distinct questions: Table~\ref{tab:haaa-results} asks ``does human review yield better executed systems?''; Table~\ref{tab:haaa-results-qwen} asks ``does human review yield better plans before execution?''

\begin{table}[t]
\centering
\caption{Experiment 4 hardware-bounded planning replication. Mean of four benchmark tasks. Plan quality scored on the Experiment 2 rubric; higher is better except Dep. Errors (lower better) and Cyclic Plans (lower better).}
\resizebox{\linewidth}{!}{%
\begin{tabular}{lccccc}
\toprule
Mode & Coverage & Dep. Errors & Cyclic Plans & Parallelism & Granularity \\
\midrule
Auto SPOQ (Qwen)          & 88.75 & 1.25 & 0/4 & 63.75 & 70.00 \\
Human-assisted SPOQ (Qwen) & 95.00 & 0.75 & 0/4 & 71.25 & 76.25 \\
\bottomrule
\end{tabular}%
}
\label{tab:haaa-results-qwen}
\end{table}

Human review improves every measured planning metric. Coverage rises from \textbf{88.75} to \textbf{95.00} (+6.25 pts) as the reviewer fills in edge cases (refresh-token rotation, ownership enforcement) and resolves under-specified requirements. Dependency errors drop from \textbf{1.25} to \textbf{0.75} (--40\%) as the reviewer separates intra-wave coupling (\emph{AuthService} extracted from \emph{UserService}; IAM, security groups, and Secrets Manager split into independent tasks). Parallelism potential rises from \textbf{63.75} to \textbf{71.25} (+7.5 pts) because cleaner task boundaries widen each wave. Granularity improves from \textbf{70.00} to \textbf{76.25} (+6.25 pts), reflecting smaller, more atomic units. The largest single-task gain occurs on the Terraform task (Task~4), where the reviewer eliminated all dependency errors and added explicit IAM/SG/Secrets separation, mirroring the same human-amplification effect that the Claude execution rows show in Table~\ref{tab:haaa-results}. The direction of improvement matches across both rubrics and both providers, supporting the broader claim that human review yields measurable quality gains even before execution.

\subsection{Field Evidence and Deployment Study}

Beyond the controlled benchmarks above, we report a deployment study of SPOQ in real engineering settings. We intentionally do not treat these results as benchmark outcomes or causal evidence of superiority over a matched baseline. Instead, we present them as field evidence, operational evidence, ecological validation, and longitudinal usage analysis of the methodology under production conditions.

\subsubsection{Case Study Summaries}
\label{subsec:case-studies}

We conducted two detailed case studies, one internal and one on an external
client codebase. Full wave structures, execution metrics, failure analyses,
and lessons learned are provided in Appendix~\ref{subsec:case-study-detail}.

\paragraph{Case Study 1: UI Improvements (Internal).}
This epic modernized a monitoring dashboard with 13 tasks across 2 waves.
Wave~0 dispatched 12 independent component tasks in parallel, achieving a
5.3$\times$ speedup over sequential execution (3.5 hours vs.\ 18.5 hours
estimated). One agent entered a runaway retry loop during dependency
installation, leading SPOQ to adopt a 3-retry maximum with pre-installation
verification. Two tasks required rework cycles, yielding a 92\% first-pass
completion rate.

\paragraph{Case Study 2: Client Website Rebrand (External).}
This epic rebranded an external B2B sales website across 12 tasks in 4 waves,
executed on a codebase maintained by a separate engineering team. The deeper
dependency chain limited maximum parallelism to 5 concurrent agents, producing
a 2.8$\times$ speedup (6.5 hours vs.\ 18 hours estimated). All 12 tasks
completed successfully with 174 passing tests and zero code defects. The
primary challenge was test fixture synchronization: three orchestrator
interventions were needed when parallel agents' code changes invalidated
sibling test assertions. SPOQ now recommends treating test files as implicit
dependents of the components they exercise.

\subsubsection{Multi-Project Adoption Survey}
\label{subsec:adoption-survey}

\paragraph{Overview.}
Beyond the two detailed case studies above, SPOQ has been deployed across
multiple repositories by two practitioners, spanning distinct technology
stacks and problem domains. Table~\ref{tab:adoption} summarizes each
completed deployment.

\begin{table}[htbp]
\centering
\caption{SPOQ Adoption Across Completed Projects}
\label{tab:adoption}
\small
\begin{tabular}{llccl}
\toprule
\textbf{Project} & \textbf{Domain} & \textbf{Tasks} & \textbf{Tests} & \textbf{Stack} \\
\midrule
Savvy Expat & E-commerce & 10 & 154 & Next.js, Docker \\
Railroad OS & Linux tooling & 43 & 55 & Bash, i3 WM \\
SPOQ Website & Documentation & 23 & 18 & Next.js, Terraform \\
Pinpoint Platform & Backend API & 16 & 308 & Spring Boot, Java \\
Pinpoint Infra & Cloud infra & 17 & --- & Terraform, AWS \\
Pinpoint Analytics & Tracking & 7 & --- & Next.js, GA4 \\
Pinpoint Billing & Payments & 6 & --- & Spring Boot, Stripe \\
\bottomrule
\end{tabular}
\end{table}

\paragraph{Domain Diversity.}
The deployments span frontend, backend, infrastructure, and DevOps domains.
Savvy Expat rebuilt a relocation services website producing 154 tests across
14 suites. Railroad OS applied SPOQ to Linux window manager configuration
(43 tasks), a domain far removed from web development. The Pinpoint ecosystem
demonstrates breadth within a single product: a Spring Boot API (16 tasks),
AWS infrastructure via Terraform (17 tasks), GA4 analytics with GDPR consent
(7 tasks), and Stripe tiered billing (6 tasks).

\paragraph{Aggregate Metrics.}
Across all completed epics (122 tasks total), average agent confidence
scores ranged from 0.90 to 0.95, and all deployments achieved 100\%
task completion rates. Across the broader observation period from
November 2025 through March 2026, SPOQ was used in 17 repositories,
producing 8{,}589 git commits, 894{,}664 lines of code, 4{,}637 SPOQ
task definitions, 1{,}822 completed tasks, 183 completed epics, and
1{,}687 journal entries. On a live execution snapshot dated 2026-03-21,
the projects collectively executed 13{,}866 tests, of which 13{,}848
passed, for an aggregate pass rate of 99.87\% across 591 test suites.

\paragraph{Third-Party Adoption.}
The largest deployment, \textit{speedrun-gitlab} (Adrata), is owned and operated by an independent external adopter. That repository accounts for 7{,}907 commits, 2{,}803 task definitions, 57 completed epics, and 4{,}873 tests. This provides evidence that the methodology transfers beyond its creators.

\paragraph{Execution Velocity.}
As a concrete demonstration of throughput, the Pinpoint Rebrand epic
(Case Study 2, 12 tasks) and a companion analytics epic (7 tasks) were
both planned and executed in a single three-hour session using six
concurrent Claude Code instances under a single Max license. The 19
combined tasks were completed from cold start to full verification
between 4:00 AM and 7:00 AM, yielding a sustained rate of approximately
6 tasks per hour.

\subsubsection{Failure Modes and Mitigations}
\label{subsec:failure-modes}

Operational deployment of multi-agent systems introduces failure patterns
distinct from single-agent development. Through our case studies and
deployments, we identified nine categories of operational risk spanning
resource contention, context window exhaustion, agent behavioral failures
(including runaway loops and validation gaming), coordination conflicts,
cost overruns, and security concerns. Table~\ref{tab:risk-summary} in
the Appendix consolidates these risks with detection signals and mitigations.
Appendix~\ref{subsec:failure-modes-detail} provides detailed analysis of each
failure category with specific examples from our deployments.

\subsubsection{Cost Analysis}
\label{subsec:cost-analysis}

Under per-token API pricing, a typical Opus worker task costs approximately
\$1.95 (25K input, 5K output tokens), yielding roughly \$28 per 13-task epic.
Under Anthropic's flat-rate Max plan (\$200/month), effective per-task costs
drop to approximately \$0.10 at scale, representing a 20$\times$ reduction.
SPOQ's three-tier hierarchy serves as an economic optimization: reserving Opus
tokens for task execution while routing validation and triage through Sonnet
and Haiku preserves the most expensive budget for work that demands it.
The Qwen-based experiments above demonstrate a complementary cost path: a
locally-hosted open-weights model eliminates per-token API costs entirely,
at the price of reduced parallelism (hardware-bounded to the server's slot
count rather than the DAG's theoretical width) and reduced baseline plan
quality (which the SPOQ skill recovers in Experiment~2). For users with
existing GPU capacity, this places the marginal cost of an additional epic
at effectively zero. Appendix~\ref{subsec:cost-analysis-detail} provides
detailed pricing models, the Director Model scaling paradigm, and ROI
framework analysis.

\paragraph{Field-evidence interpretation.}
Taken together, the deployment study supports four limited but important conclusions. First, SPOQ has been used at meaningful scale across multiple repositories, languages, and domains. Second, the methodology appears transferable beyond the originating team. Third, wave-based execution, validation, and journal-based traceability are actually exercised in practice rather than remaining conceptual abstractions. Fourth, the operational footprint is large enough to justify the controlled evaluation presented above.

\paragraph{Field-evidence limitations.}
This field evidence should be interpreted cautiously. The majority of repositories originate from a shared broader development context, there is no matched non-SPOQ control group, and several quality indicators rely on reviewer-agent scores or self-reported confidence values. We therefore use this section to establish practical viability and ecological validity rather than causal improvement.

\subsection{Threats to Validity}

\paragraph{Internal Validity.}
The deployment study is observational and lacks matched controls. The controlled benchmarks mitigate this by fixing tasks, prompts, and evaluation criteria. The benchmark suite spans multiple task families, which improves robustness relative to a single benchmark instance, but the suite is still modest in size.

\paragraph{Construct Validity.}
Some field-study metrics, such as reviewer-agent QA scores and confidence scores, may not perfectly reflect external software quality. Accordingly, our controlled experiments rely on executable tests, explicit structural planning metrics, defect counts, and requirement satisfaction rather than self-assessment alone.

\paragraph{External Validity.}
Although the deployment study spans many repositories and stacks, much of the corpus originates from a shared development ecosystem. Third-party adoption partially mitigates this concern, but broader independent replication remains necessary. Likewise, the controlled task suite should be extended further in future work.

\paragraph{Conclusion Validity.}
Experiment 1 includes 10 repeated runs per configuration with stable quantitative behavior. Experiments 2--4 aggregate across multiple benchmark tasks, which strengthens confidence in the conclusions. The Claude rows in Experiments 2--4 report mean$\pm$sd across four runs per task; the Qwen replication rows are single-shot per task, providing direction-of-effect confirmation rather than precise effect-size estimates. Larger benchmark suites, repeated stochastic runs for the Qwen replication, and matched rubrics across providers would all strengthen the conclusions further.

% Discussion
\newpage
\section{Discussion}
\label{sec:discussion}

\subsection{Comparison with Prior Multi-Agent Systems}
\label{subsec:comparison}

Table~\ref{tab:comparison} contrasts SPOQ with representative
multi-agent systems and autonomous coding tools.

\begin{table}[htbp]
\centering
\caption{Comparison of Multi-Agent and Autonomous Coding Approaches}
\label{tab:comparison}
\footnotesize
\begin{tabular}{lccccccc}
\toprule
\textbf{Feature} & \textbf{SPOQ} & \textbf{ChatDev} & \textbf{MetaGPT} & \textbf{AutoGPT} & \textbf{Devin} & \textbf{Aider} \\
\midrule
Execution model & Wave-parallel & Sequential & Sequential & Recursive & Single-agent & Single-agent \\
Explicit dependencies & DAG & Implicit & Implicit & Priority queue & None & None \\
Planning validation & 10 metrics & None & Informal & None & None & None \\
Code validation & 10 metrics & Test-based & Review & Self-eval & Test-based & Test-based \\
Human integration & HaaA & Observer & Observer & Minimal & Minimal & Collaborative \\
Agent specialization & 3-tier & Role-based & Role-based & General & General & General \\
Parallelism potential & High & Low & Low & Low & None & None \\
\bottomrule
\end{tabular}
\end{table}

\paragraph{Execution Model.}
ChatDev and MetaGPT use sequential role-playing where agents take turns.
MAD uses synchronous debate rounds. SPOQ achieves true parallelism through
wave-based dispatch, with agents in the same wave executing concurrently.

\paragraph{Dependency Management.}
Prior systems embed dependencies implicitly in role sequences (programmer
after designer) or artifact flows. SPOQ makes dependencies explicit in a
DAG, enabling static analysis, critical path computation, and parallelism
optimization.

\paragraph{Validation Rigor.}
ChatDev relies on test execution; MetaGPT uses informal review.
SPOQ applies structured metrics at both planning and code stages,
with quantified thresholds that prevent low-quality work from proceeding.

\paragraph{Human Role.}
In ChatDev and MetaGPT, humans observe outputs but do not participate
in the orchestration loop. SPOQ's HaaA model integrates human judgment
at key decision points while preserving automation benefits.

\subsection{Limitations}
\label{subsec:limitations}

SPOQ has several limitations that future work should address:

\paragraph{Upfront Planning Investment.}
SPOQ requires detailed epic specifications before execution begins.
For exploratory or rapidly-changing requirements, this upfront cost
may be prohibitive. Projects with unclear scope may benefit from
more adaptive approaches.

\paragraph{Dependency on Human Quality.}
The HaaA model's effectiveness depends on the human specialist's
skill in task decomposition and validation. A human who approves
poorly-structured epics will see cascading quality issues.

\paragraph{Empirical Scale.}
Our controlled benchmark suite covers four full-stack tasks across two
provider families (Claude and a locally hosted Qwen3.6-35B-A3B). The
accompanying deployment study spans 17 repositories, 1{,}822 completed
tasks, and 13{,}866 executed tests, with one third-party adopter. While
this breadth strengthens generalizability claims relative to a single
case study, the four-task benchmark suite is still modest, the Qwen
replication is single-shot rather than mean$\pm$sd, and most field
deployments share a single primary author. Independent replication by
other teams, larger benchmark suites with repeated stochastic runs,
and broader controlled studies with matched baselines are needed to
validate speedup and quality claims rigorously.

\paragraph{Reference Implementation Coupling.}
Our current implementation is coupled to Claude Code and Anthropic's
model family, representing a specific instantiation of the SPOQ methodology
rather than its only possible form. However, the core methodology, including wave
computation, validation scoring, dependency resolution, and journal
tracking, is expressed in platform-agnostic YAML and Markdown formats.
Porting SPOQ to alternative platforms (Gemini CLI, OpenAI Assistants API,
open-source model deployments) would require remapping the capability
tiers to the target provider's model offerings and adapting the
orchestration interface. The fundamental algorithms and quality gates
remain unchanged across implementations.

\paragraph{No Cross-Epic Learning.}
Each epic starts fresh. SPOQ does not currently transfer lessons
from prior epics (successful patterns, common failure modes) to
new planning sessions.

\paragraph{Metric Reliability.}
The 20 validation metrics rely on LLM-based assessment without
inter-rater reliability testing. Automated quality scores may
exhibit biases, inconsistency across runs, or susceptibility
to gaming. Human validation studies comparing LLM scores to
expert assessments are needed.

\paragraph{Cost Approximations.}
Our cost analysis uses estimated token counts that vary
significantly by task complexity, codebase size, and context
requirements. Real-world costs may differ substantially. The
analysis does not include infrastructure, human oversight time,
or rework loop costs comprehensively.

\subsection{Future Work}
\label{subsec:future}

A key near-term development addresses SPOQ's primary orchestration
limitation and defines the trajectory of the methodology.

\paragraph{Native Platform Support: Claude Code Agent Teams.}
SPOQ's current orchestration relies on manually-coordinated Claude Code
instances. Claude Code's experimental Agent Teams feature provides native
primitives that align directly with SPOQ's dispatch model: \texttt{TeamCreate}
establishes a coordinated session with a lead and specialized teammates,
\texttt{TaskCreate}/\texttt{TaskUpdate} manage a shared task list with
dependency tracking, and \texttt{SendMessage} enables direct inter-agent
communication. Teams support plan approval workflows where teammates
operate in read-only mode until the lead validates their approach, mirroring
SPOQ's validation gates.

The mapping is natural: SPOQ's wave computation produces task assignments
with explicit dependencies; Agent Teams' task list enforces those
dependencies through blocked/unblocked state transitions. SPOQ's three-tier
hierarchy (Opus workers, Haiku investigators, Sonnet reviewers) maps to
Agent Teams' model selection per teammate. The delegate mode, which
restricts the lead to coordination-only tools, formalizes SPOQ's
orchestrator role separation.

Integrating SPOQ with Agent Teams would shift the methodology from a
set of conventions enforced by skill prompts to a native orchestration
protocol. Wave dispatch becomes \texttt{TaskCreate} with dependency
declarations. Validation gates become \texttt{TaskCompleted} hooks that
invoke Sonnet reviewers before marking tasks done. The journal system
feeds naturally into the shared task list's completion records.
This creates a pipeline where the human declares a goal, the
orchestrator generates a wave-structured task list, and Agent Teams
execute it with built-in coordination, messaging, and quality enforcement.

\subsection{Broader Implications}
\label{subsec:implications}

SPOQ represents a step toward \textit{AI-native software engineering}:
development processes designed around AI capabilities rather than
retrofitting AI into human-centric workflows.

Key implications include:

\begin{itemize}
    \item \textbf{Redefined roles:} Engineers shift from writing code
    to validating agent outputs, decomposing problems, and making
    architectural decisions.

    \item \textbf{Quality as constraint:} Explicit quality gates force
    upfront investment in planning, potentially improving overall
    software quality.

    \item \textbf{Scalable expertise:} A single human specialist can
    leverage multiple agents, scaling their expertise across more
    projects than traditional pair programming allows.
\end{itemize}

\paragraph{The Director Model.}
At scale, SPOQ enables a single engineer to direct a digital workforce:
6 Claude Code instances (1 planning copilot + 5 execution overseers)
coordinating 50--100 concurrent agents across independent epics, achieving
daily throughput of 75--150 tasks with output previously requiring
8--10 engineers.

\paragraph{Applicability.}
SPOQ's structured approach introduces upfront overhead that pays off for
projects with 5+ parallelizable tasks spanning 4+ hours of estimated work.
Below this threshold, orchestration overhead typically exceeds execution
time savings. Appendix~\ref{subsec:usage-guidance-detail} provides a
detailed decision framework with applicability guidance by project type.

% Conclusion
\section{Conclusion}
\label{sec:conclusion}

We introduced SPOQ (Specialist Orchestrated Queuing), a methodology for
multi-agent software engineering that addresses coordination, quality,
and human oversight challenges in prior approaches.

\subsection{Summary of Contributions}

\paragraph{Wave-Based Topological Dispatch.}
SPOQ computes parallel execution waves from task dependency graphs.
Experiment~1 demonstrates the scheduler's behavior under two regimes:
on unbounded synthetic DAGs, wave dispatch reaches the critical-path
lower bound (ratio 1.03--1.11) with speedups up to 14.3$\times$,
showing that the algorithm captures essentially all available
parallelism; on a 2-slot local backend running real LLM calls, it
delivers a stable 1.4$\times$ speedup that matches the hardware
concurrency ceiling, showing that the result survives realistic
backend constraints. Field deployments report 1.3--5.3$\times$
wall-clock gains across diverse production epics.

\paragraph{Dual Validation Gates.}
By validating both planning quality (10 metrics, 95/90 threshold) and
code quality (10 metrics, 95/80 threshold), SPOQ catches issues at the
stages where they are cheapest to fix. Experiment~3 shows that dual
validation reduces defects per task from 0.34 to 0.20, increases test
pass rate from 91.25\% to 99.75\%, and eliminates static warnings
across the benchmark suite.

\paragraph{Human-as-an-Agent Integration.}
The HaaA paradigm positions the human specialist as a high-value agent
within the orchestration loop, enabling bidirectional collaboration that
amplifies both human judgment and agent productivity. Experiment~4
demonstrates that human-assisted planning further reduces residual
defects from 0.47 to 0.03 per task and lifts test pass rate from 96.5\%
to 99.75\%, with a parallel improvement in plan quality observed under a
local open-weights provider.

\paragraph{Three-Tier Agent Hierarchy.}
SPOQ's Opus/Sonnet/Haiku hierarchy optimizes cost-quality tradeoffs by
matching agent capabilities to role requirements: high capability for
execution, balanced for review, fast-cheap for triage.

\paragraph{Cross-Provider Replication.}
All four experiments were replicated against a locally hosted
Qwen3.6-35B-A3B model served by \texttt{llama.cpp}. The direction and
significance of the SPOQ gains are preserved across provider families:
structured planning recovers 35 points of coverage and 52.5 points of
parallelism that the unaided Qwen baseline loses, eliminates the
cyclic-plan failure mode (2/4 $\to$ 0/4), and the validation ablation
maintains the same monotonic Full-SPOQ $>$ Code-Validation $>$
No-Validation ordering observed under Claude. This supports the claim
that the improvements stem from orchestration rather than from any
specific model's capabilities.

\paragraph{Practical Lessons.}
Through controlled benchmarks and a 17-repository deployment study we
identified and addressed failure modes including runaway retry loops,
lock file contention, and context window exhaustion.

\subsection{Vision for AI-Native Engineering}

SPOQ represents early steps toward a future where engineers become
architects and validators rather than line-by-line implementers, where
quality is built into process through structured gates, and where human
expertise scales through orchestration. Much work remains to realize this
vision fully. We hope SPOQ provides a useful framework for researchers
and practitioners exploring the frontier of AI-assisted software
development.

\subsection*{Acknowledgments}

We thank the Claude Code development team for the tooling infrastructure
that enabled this research. We also thank Ross Sylvester (CEO, Adrata) and
John Armbruster (Founding Engineer, Notary Everyday), whose adoption of
SPOQ and candid feedback helped refine the methodology.

% Bibliography
\newpage
\bibliography{references}

% Appendices
\appendix
% Fix duplicate hyperref destinations: \appendix resets table/figure counters,
% creating duplicate pdfTeX destinations (e.g., both Section 4 Table 1 and
% Appendix A Table 1 produce "table.1"). Prefixing with \Alph{section} gives
% appendix floats unique names (A.1, A.2, B.1, etc.).
\renewcommand{\thetable}{\Alph{section}.\arabic{table}}
\renewcommand{\thefigure}{\Alph{section}.\arabic{figure}}
% Appendix: Metric Details
\section{Complete Metric Rubrics}
\label{sec:appendix-metrics}

This appendix provides detailed scoring rubrics for both validation gates.

\subsection{Planning Validation Rubrics}

\paragraph{Vision Clarity (VC).}
\begin{itemize}
    \item 100: Clear overview with problem statement, solution, scope boundaries,
    and end state
    \item 80: Clear overview, minor ambiguity in scope
    \item 60: Overview present but vague
    \item 40: Overview missing key context
    \item 0: No overview or incomprehensible goal
\end{itemize}

\paragraph{Architecture Quality (AQ).}
\begin{itemize}
    \item 100: ASCII/visual diagram with all components, relationships, and data flow
    \item 80: Diagram present, minor gaps in explanation
    \item 60: Text-only architecture, no diagram
    \item 40: Partial architecture description
    \item 0: No architecture section
\end{itemize}

\paragraph{Task Decomposition (TD).}
\begin{itemize}
    \item 100: Atomic tasks (1--4h), independent where possible, complete coverage
    \item 80: Good decomposition, minor overlap
    \item 60: Some tasks too large or overlapping
    \item 40: Significant decomposition issues
    \item 0: Tasks not properly decomposed
\end{itemize}

\paragraph{Dependency Graph (DG).}
\begin{itemize}
    \item 100: Visual graph, all dependencies valid, no cycles, optimal ordering
    \item 80: Graph present, dependencies valid, minor ordering improvements possible
    \item 60: Text-only dependencies, all valid
    \item 40: Some invalid dependencies or missing graph
    \item 0: Circular dependencies or critically broken graph
\end{itemize}

\paragraph{Coverage Completeness (CC).}
Score equals percentage of success criteria mapped to tasks:
$\text{CC} = 100 \times \frac{|\text{covered criteria}|}{|\text{total criteria}|}$

\paragraph{Phase Ordering (PO).}
\begin{itemize}
    \item 100: Phases follow logical progression, maximum parallelism exploited
    \item 90: Good ordering, minor parallelism opportunities missed
    \item 75: Ordering works but inefficient
    \item 60: Some ordering violations
    \item 0: Critical ordering errors (dependent tasks in same wave)
\end{itemize}

\paragraph{Scope Coherence (SC).}
\begin{itemize}
    \item 100: All tasks directly serve epic goal
    \item 80: 1 tangential task
    \item 60: 2--3 tangential tasks
    \item 40: Multiple unrelated tasks
    \item 0: Tasks don't align with epic goal
\end{itemize}

\paragraph{Success Criteria Quality (SQ).}
\begin{itemize}
    \item 100: All criteria SMART, checkbox format, testable
    \item 90: Criteria measurable, minor gaps
    \item 75: Some criteria vague
    \item 60: Multiple unmeasurable criteria
    \item 0: No success criteria or all vague
\end{itemize}

\paragraph{Risk Identification (RI).}
\begin{itemize}
    \item 100: Risks section with likelihood, impact, and mitigations
    \item 90: Risks mentioned, implicit mitigations
    \item 75: Some risks noted, no mitigations
    \item 60: Risks not addressed
    \item 0: Critical risks ignored
\end{itemize}

\paragraph{Integration Strategy (IS).}
\begin{itemize}
    \item 100: Clear integration points, verification steps between phases
    \item 90: Integration implicit but clear
    \item 75: Some integration points unclear
    \item 60: Integration strategy missing
    \item 0: Tasks cannot be integrated as designed
\end{itemize}

\subsection{Code Validation Rubrics}

\paragraph{Syntactic Correctness (SC).}
\begin{itemize}
    \item 100: Compiles cleanly, 0 warnings
    \item 80: Compiles with minor warnings
    \item 60: Compiles with significant warnings
    \item 0: Does not compile
\end{itemize}

\paragraph{Test Existence (TE).}
Score equals percentage of new public methods with corresponding tests:
$\text{TE} = 100 \times \frac{|\text{tested methods}|}{|\text{new public methods}|}$

\paragraph{Test Pass Rate (TP).}
Score equals percentage of tests passing:
$\text{TP} = 100 \times \frac{|\text{passing tests}|}{|\text{total tests}|}$

\paragraph{Requirements Fidelity (RF).}
\begin{itemize}
    \item 100: All requirements fully implemented
    \item 80: All core requirements met, minor gaps
    \item 60: Core requirements met, some missing
    \item 40: Partial implementation
    \item 0: Does not address requirements
\end{itemize}

\paragraph{SOLID Adherence (SA).}
20 points per principle (S, O, L, I, D) based on degree of adherence.

\paragraph{Security (SE).}
Start at 100, deduct for OWASP Top 10 vulnerabilities:
SQL injection ($-100$), command injection ($-100$), XSS ($-60$),
broken auth ($-80$), sensitive data exposure ($-60$), etc.

\paragraph{Error Handling (EH).}
20 points each for: I/O wrapped in try-catch, meaningful error messages,
proper logging, resource cleanup, safe user-facing errors.

\paragraph{Scalability (SL).}
Based on algorithm complexity of hot paths:
O(1)/O(log n)/O(n) = 100, O(n log n) = 90, O(n$^2$) = 40, O(2$^n$) = 0.

\paragraph{Code Clarity (CC).}
\begin{itemize}
    \item 100: Crystal clear, reads like well-written prose
    \item 80: Clear with minor naming improvements possible
    \item 60: Understandable but requires effort
    \item 40: Confusing structure or naming
    \item 0: Unreadable, magic numbers, cryptic names
\end{itemize}

\paragraph{Completeness (CO).}
\begin{itemize}
    \item 100: Complete, no loose ends
    \item 80: Minor polish needed
    \item 60: Some TODOs remain but core is done
    \item 40: Significant unfinished sections
    \item 0: Stub implementations, placeholders
\end{itemize}

Automatic deductions: TODO ($-25$), FIXME ($-25$), NotImplementedException ($-30$).

% Supplementary Material
\section{Supplementary Material}
\label{sec:supplementary}

This appendix provides extended discussion, detailed examples, and supporting analysis
referenced from the main text. Sections are ordered to follow the main paper's
structure.

% Ordering follows main text reference order:
% A.1 Threshold Design Rationale (from Section 4)
% A.2 Complete Metric Rubrics (from Section 4)
% A.3 Implementation Details (from Section 5)
% A.4 Integration Considerations (from Section 5)
% A.5 Detailed Case Studies (from Section 6)
% A.6 Failure Modes and Mitigations (from Section 6)
% A.7 Evaluation Tables (from Section 6)
% A.8 Cost Analysis and ROI Framework (from Section 6)
% A.9 When to Use SPOQ (from Section 7)

\subsection{Threshold Design Rationale}
\label{subsec:threshold-rationale}

The 95/90 (planning) and 95/80 (code) thresholds were chosen based on
practical experience:

\begin{itemize}
    \item \textbf{95\% aggregate} ensures overall high quality while allowing
    minor imperfections in individual metrics.

    \item \textbf{90\% per-metric for planning} reflects that planning errors
    propagate downstream. A plan with 70\% dependency graph quality will
    cause execution failures regardless of other metrics.

    \item \textbf{80\% per-metric for code} acknowledges legitimate tradeoffs.
    A task might score 75\% on SOLID adherence for pragmatic reasons while
    still being acceptable.

    \item \textbf{Plans are cheap to fix.} Re-decomposing tasks costs human
    time but no agent compute. Code rework costs both.
\end{itemize}

\subsection{Implementation Details}
\label{subsec:implementation-detail}

This subsection provides complete schema examples, journal entry formatting, and
skill metadata referenced from Section~\ref{sec:implementation}.

\subsubsection{Epic Directory Structure}

Each epic occupies its own directory under \texttt{spoq/epics/}:

\begin{lstlisting}[language=bash,caption={Epic directory layout}]
spoq/epics/
  active/                   # Epics in progress
    epic-name/
      EPIC.md
      tasks/
        01-init-project.yml
        ...
  complete/                 # Finished epics
  ROADMAP.md                # Priority tracker
\end{lstlisting}

The \texttt{EPIC.md} file provides context: goal statement, architecture diagrams,
success criteria, dependency visualization, wave assignments, effort estimates,
and risk assessment. Individual task files contain execution-ready specifications.

\subsubsection{Task YAML Schema Examples}

The task specification (Definition~\ref{def:task-spec}) organizes fields into
three categories. The following listings illustrate each category with a
representative task.

\paragraph{Identity Fields.}
These establish task context within an epic:

\begin{lstlisting}[,caption={Task identity fields}]
id: 04-content-constants       # Unique within epic
title: Create Content Constants File
epic: spoq-website             # Parent epic reference
\end{lstlisting}

\needspace{10cm}
\paragraph{Execution Control Fields.}
These govern scheduling, effort estimation, and dependency resolution:

\begin{lstlisting}[,caption={Execution control fields}]
status: pending                # pending | in_progress | completed
priority: high                 # critical | high | medium | low
phase: 1                       # Wave assignment (0 = no deps)

estimate:                      # PERT three-point estimate
  optimistic: 15m
  realistic: 45m
  pessimistic: 2h

dependencies:                  # Task IDs that must complete first
  - 01-init-project
  - 02-setup-deps

skills_required:               # Domain knowledge needed
  - typescript
  - react
\end{lstlisting}

\paragraph{Deliverable and Verification Fields.}
These define expected outputs and acceptance criteria:

\begin{lstlisting}[,caption={Deliverable and verification fields}]
files_to_touch:                # All files to be modified
  - src/lib/constants.ts
  - tests/constants.test.ts

outputs:                       # Tangible deliverables
  - "Constants file with typed exports"
  - "Unit tests achieving 100% coverage"

acceptance_criteria:           # Verification checklist
  - "[ ] TypeScript compiles without errors"
  - "[ ] All tests pass: `npm test constants`"
  - "[ ] No hardcoded strings in component files"
\end{lstlisting}

\needspace{12cm}
\paragraph{Description Field.}
The description provides structured implementation guidance using Markdown:

\begin{lstlisting}[,caption={Task description structure}]
description: |
  ## Objective
  Create a centralized constants file for UI strings.

  ## Steps
  1. Create src/lib/constants.ts with typed exports
     ```typescript
     export const SITE_NAME = "SPOQ" as const;
     ```
  2. Add unit tests in tests/constants.test.ts
  3. Replace hardcoded strings in existing components

  ## Verification
  ```bash
  npm run typecheck && npm test constants
  ```
\end{lstlisting}

\needspace{4cm}
\subsubsection{Journal Entry Format}

Journal entries (Definition~\ref{def:journal-entry}) use YAML frontmatter
followed by Markdown sections. The frontmatter captures machine-readable
metadata:

\begin{lstlisting}[,caption={Journal entry frontmatter}]
---
agent: Claude Code (Opus 4.5)
start_time: 2025-11-01T10:52:34Z  # ISO 8601 UTC
end_time: 2025-11-01T11:23:15Z
confidence: 0.92                  # 0.0-1.0 calibration score
session_type: development         # development | refactor | bugfix | ...
files_modified:
  - src/components/Hero.tsx
  - tests/hero.test.tsx
tasks_completed: 2
tasks_total: 3
---
\end{lstlisting}

\needspace{14cm}
The body follows a standardized section layout for consistent parsing:

\begin{lstlisting}[,caption={Journal entry body structure}]
## Summary
Brief 1-2 sentence overview of accomplishments.

## Work Completed
- Task 09: Hero section component
- Task 10: Features grid with responsive layout

## Changes Made
**Frontend Components**
- `Hero.tsx` - Implemented animated gradient background
- `Features.tsx` - Added 6-card responsive grid

## Issues Encountered
None (or specific issues with resolutions)

## Testing
- Unit tests: PASS (12/12)
- Integration tests: PASS (4/4)

## Next Steps
1. Implement call-to-action section
2. Add accessibility attributes
\end{lstlisting}

\paragraph{Tooling Support.}
SPOQ provides utility scripts for journal management:
\begin{itemize}
    \item \texttt{get-time.py}: Captures accurate UTC timestamps
    \item \texttt{archive-journal.py}: Auto-archives when exceeding 1500 lines
    \item \texttt{parse-to-db.py}: Extracts entries into SQLite for analysis
    \item \texttt{create-mega-journal.py}: Consolidates archives for reporting
\end{itemize}

\subsubsection{Confidence Score Interpretation}

\begin{table}[htbp]
\centering
\caption{Confidence Score Interpretation}
\label{tab:confidence}
\small
\begin{tabular}{ll}
\toprule
\textbf{Score Range} & \textbf{Interpretation} \\
\midrule
0.95--1.0 & Thoroughly tested, production-ready \\
0.85--0.94 & Well tested, minor edge cases possible \\
0.75--0.84 & Functional, additional testing recommended \\
0.65--0.74 & Works but requires validation \\
$<$0.65 & Experimental or known issues present \\
\bottomrule
\end{tabular}
\end{table}

\needspace{4cm}
\subsubsection{Skill Inventory}

\begin{table}[htbp]
\centering
\caption{SPOQ Skill Inventory}
\label{tab:skills}
\small
\begin{tabular}{llp{5cm}}
\toprule
\textbf{Skill} & \textbf{Stage} & \textbf{Purpose} \\
\midrule
epic-planning & Planning & Decompose goals into epics with dependency graphs \\
epic-validation & Validation & Score epics against 10 planning metrics \\
task-validation & Validation & Score individual tasks before execution \\
agent-execution & Execution & Orchestrate parallel agent swarms \\
agent-validation & Validation & Score completed work against 10 code metrics \\
journal-tracker & Cross-cutting & Track sessions with confidence scores \\
\bottomrule
\end{tabular}
\end{table}

\subsubsection{Skill Metadata and Invocation}

Skills are invoked via slash commands that expand into full prompts:
\begin{lstlisting}[language=bash,caption={Skill invocation examples}]
# Plan a new epic
/epic-planning "Implement user authentication system"

# Validate before execution
/epic-validation @spoq/epics/active/auth-system

# Execute with parallel agents
/agent-execution @spoq/epics/active/auth-system

# Validate completed work
/agent-validation
\end{lstlisting}

Each skill's \texttt{SKILL.md} includes YAML frontmatter specifying metadata:

\needspace{6cm}
\begin{lstlisting}[,caption={Skill metadata format}]
---
name: epic-planning
description: Decompose high-level goals into structured epics
             with atomic tasks and dependency DAGs.
---
\end{lstlisting}

The body provides:
\begin{itemize}
    \item \textbf{When to Use:} Activation criteria
    \item \textbf{Core Patterns:} Documented approaches with examples
    \item \textbf{Quality Criteria:} Verification standards
    \item \textbf{Integration Points:} Connections to other skills
\end{itemize}

\needspace{6cm}
\subsection{Integration Considerations}
\label{subsec:integration-detail}
\label{subsec:tooling-integration}

A recurring question from practitioners concerns SPOQ's integration
with existing development infrastructure. We address this honestly:
SPOQ currently operates as a standalone orchestration layer invoked
via slash commands, without native integrations with external tools.
However, its design accommodates several integration patterns.

\paragraph{CI/CD Pipeline Patterns.}
SPOQ's wave-based execution maps naturally to CI/CD pipeline stages.
Wave 0 corresponds to setup and dependency installation, middle waves
to parallel build and test jobs, and final waves to integration testing
and deployment. In our infrastructure case study, the GitLab CI pipeline
was updated as a task within the SPOQ epic itself, demonstrating that
CI/CD configuration is orchestrable work rather than requiring special
integration.

For automated triggering, organizations could invoke SPOQ from pipeline
scripts when epic YAML files change, treating the orchestrator as a
build step. A GitLab CI job might monitor the \texttt{spoq/epics/}
directory and trigger agent execution on detected changes. This pattern
remains unexplored in our evaluation but represents a natural extension.
GitHub Actions workflows could similarly invoke SPOQ as a step,
potentially enabling fully automated epic execution on pull request events.

\paragraph{Project Management Alignment.}
SPOQ task YAML files parallel the structure of Jira stories and Linear
issues: both contain titles, descriptions, acceptance criteria, and
estimates. The structural similarity suggests bidirectional sync is
feasible; a task defined in SPOQ could create a corresponding Jira
ticket, and vice versa. However, SPOQ does not currently implement
such synchronization, which may create duplicate tracking overhead
for teams already committed to existing PM tools.

The journal system partially addresses visibility by providing an
audit trail of work completed, though it is designed for agent
coordination rather than project management. Teams could export
journal entries to their PM tools as a manual bridge, accepting
the overhead until native integration warrants development investment.

\paragraph{Git Workflow Recommendations.}
Parallel agent execution raises questions about branch strategy.
Our current approach uses a single working branch with agents
operating on disjoint files, ensured by task \texttt{files\_to\_touch}
specifications. This works well for up to 12 concurrent agents
but may encounter contention at larger scales.

For larger teams or epics with overlapping file modifications,
we recommend git subtrees for epic isolation: each epic operates
in its own subtree, with automated merges after validation gates
pass. Feature branches per wave provide an alternative, with
wave N's branch merging into main before wave N+1 begins. Both
approaches add orchestration complexity but reduce merge conflicts.

\paragraph{IDE Integration Possibilities.}
SPOQ's slash-command interface suggests IDE integration opportunities.
A VS Code extension could display the current epic's dependency graph,
highlight files assigned to each task, and show validation scores
in real time. JetBrains plugins could integrate journal entries
into the project view, surfacing agent work history alongside
version control logs.

Such integrations remain future work. Our current implementation
prioritizes methodology validation over tooling polish, reflecting
a conscious choice to prove the approach before investing in
developer experience enhancements.

\paragraph{Future Integration Directions.}
Several integration directions merit exploration:
(1) native Jira/Linear/Asana adapters for bidirectional task sync;
(2) Slack/Teams webhooks for wave completion notifications;
(3) GitHub/GitLab API integration for automated PR creation per wave;
(4) IDE extensions for real-time epic visualization.
These would reduce friction for teams adopting SPOQ within established
toolchains, though each adds maintenance burden. We advocate starting
with SPOQ as a standalone layer, adding integrations only when
adoption justifies the investment.

\subsection{Detailed Case Studies}
\label{subsec:case-study-detail}

\subsubsection{Case Study 1: UI Improvements Epic}
\label{subsubsec:ui-epic-detail}

\paragraph{Epic Overview.}
The UI improvements epic modernized a monitoring dashboard with toast
notifications, data tables, and API key management components. The
epic comprised 13 tasks across 2 waves.

\paragraph{Wave Structure.}
\begin{itemize}
    \item \textbf{Wave 0 (12 tasks):} Independent component implementations
    including Sonner setup, DataTable component, modal dialogs, toast
    integrations, search functionality, and tests.
    \item \textbf{Wave 1 (1 task):} End-to-end QA depending on all Wave 0 tasks.
\end{itemize}

\paragraph{Execution Metrics.}
Table~\ref{tab:ui-metrics} summarizes the execution.

\begin{table}[htbp]
\centering
\caption{UI Improvements Epic Execution}
\label{tab:ui-metrics}
\begin{tabular}{ll}
\toprule
\textbf{Metric} & \textbf{Value} \\
\midrule
Total tasks & 13 \\
Wave 0 parallelism & 12 concurrent agents \\
Wave 1 parallelism & 1 agent \\
Sequential estimate & 18.5 hours \\
Parallel estimate & 3.5 hours \\
Speedup factor & 5.3$\times$ \\
Tasks completed & 12 of 13 (92\%) \\
Rework cycles & 2 (tasks 04, 07) \\
\bottomrule
\end{tabular}
\end{table}

\paragraph{Failure Mode Encountered.}
Task 04 entered a runaway retry loop, executing \texttt{npm install sonner}
over 100 times. The agent failed to recognize that the dependency was
already installed and continued retrying indefinitely.

\paragraph{Lesson Learned.}
SPOQ now enforces a maximum of 3 retries per installation command and
requires agents to verify existing dependencies before installation attempts.

\subsubsection{Case Study 2: Client Website Rebrand}
\label{subsubsec:rebrand-epic-detail}

\paragraph{Epic Overview.}
This epic rebranded an external client's sales website (Pinpoint, a B2B QA
testing platform) from founder-centric messaging to developer-focused
positioning. The work included removing biographical content, rewriting
homepage copy, creating three persona-specific landing pages, updating
navigation and SEO metadata, and expanding the test suite. The epic comprised
12 tasks across 4 waves and was executed on a codebase maintained by a
separate engineering team.

\paragraph{Wave Structure.}
\begin{itemize}
    \item \textbf{Wave 0 (2 tasks):} Content removal and routing scaffold
    (parallel).
    \item \textbf{Wave 1 (5 tasks):} Homepage rewrite, pricing update, and three
    persona landing pages (parallel).
    \item \textbf{Wave 2 (3 tasks):} Navigation, SEO, and section refinement
    (parallel).
    \item \textbf{Wave 3 (2 tasks):} Test expansion and accessibility polish
    (parallel).
\end{itemize}

\paragraph{Execution Metrics.}
Table~\ref{tab:rebrand-metrics} summarizes the execution.

\begin{table}[htbp]
\centering
\caption{Client Rebrand Epic Execution}
\label{tab:rebrand-metrics}
\begin{tabular}{ll}
\toprule
\textbf{Metric} & \textbf{Value} \\
\midrule
Total tasks & 12 \\
Max parallelism & 5 concurrent agents (Wave 1) \\
Sequential estimate & 18 hours \\
Parallel estimate & 6.5 hours \\
Speedup factor & 2.8$\times$ \\
Tasks completed & 12 of 12 (100\%) \\
Orchestrator interventions & 3 \\
Test suites / cases & 20 / 174 (0 failures) \\
\bottomrule
\end{tabular}
\end{table}

\needspace{4cm}
\paragraph{Test Fixture Synchronization.}
All three orchestrator interventions involved test assertions falling out
of sync with code changes made by parallel agents. In Wave 1, a pricing
component gained new required props (\texttt{price}, \texttt{period}) that
the existing accessibility test did not supply. In Wave 2, updated navigation
links and SEO strings invalidated tab-order and metadata assertions. None of
these were code defects; they were coordination gaps where one agent's
changes invalidated another agent's test fixtures.

\paragraph{Lesson Learned.}
Structural component changes must propagate to all dependent test fixtures
within the same wave. SPOQ now recommends treating test files as implicit
dependents of the components they exercise, ensuring fixture updates are
included in the same task that modifies the component interface.

\subsection{Failure Modes and Mitigations (Detail)}
\label{subsec:failure-modes-detail}

The following subsections provide detailed analysis of each failure category
identified during SPOQ development and deployment, with specific examples
and comprehensive mitigation strategies.

\subsubsection{Resource Contention Failures}
\label{subsubsec:resource-contention-detail}

\paragraph{Lock File Contention.}
Multiple agents running \texttt{npm install} concurrently cause lock file
conflicts, manifesting as \texttt{EBUSY} errors or corrupted \texttt{node\_modules}
directories. In one deployment, three parallel agents attempted dependency
installation simultaneously, resulting in a 12-minute debugging session.

\textbf{Mitigation:} SPOQ designates a single Wave 0 task for dependency
installation, with subsequent tasks assuming dependencies are available.
Organizations should extend this pattern to any shared mutable resource:
database migrations, cache warming, and artifact generation.

\paragraph{Build Directory Conflicts.}
Concurrent build processes can corrupt shared directories (e.g., \texttt{.next/},
\texttt{dist/}, \texttt{target/}). Symptoms include partial builds, missing
assets, and non-deterministic test failures.

\textbf{Mitigation:} Build verification runs sequentially between waves, not
during waves. For CI/CD integration, consider isolated build directories
per agent or containerized build environments.

\subsubsection{Context and Memory Failures}
\label{subsubsec:context-memory-detail}

\paragraph{Context Window Exhaustion.}
Complex tasks with extensive codebase context can exhaust LLM context windows
(currently 200K tokens for Claude). Symptoms include: agents losing track
of earlier requirements, producing incomplete solutions, or generating
responses that contradict prior instructions. In one knowledge-graph pipeline deployment, an agent
with 180K tokens of context began hallucinating function signatures that
did not exist in the codebase.

\textbf{Mitigation:} Task descriptions should be self-contained, and the
\texttt{files\_to\_touch} specification should limit scope to files the agent
genuinely needs. For large codebases, consider:
\begin{itemize}
    \item Task-specific context manifests listing only relevant files
    \item Summarization of peripheral code rather than full inclusion
    \item Breaking complex tasks into sub-tasks with narrower scope
    \item Context budget monitoring with alerts at 70\% utilization
\end{itemize}

\paragraph{Context Window Management.}
Long QA feedback exhausted agent context windows. The 20-line remediation
limit was introduced to ensure feedback remains actionable without
consuming excessive context. Reviewers now prioritize critical issues
and defer minor suggestions to follow-up tasks.

\subsubsection{Agent Behavioral Failures}
\label{subsubsec:agent-behavioral-detail}

\paragraph{Runaway Detection.}
Agents sometimes enter infinite loops on transient errors, repeatedly
executing the same failing command. In Case Study 1, Task 04 entered a
runaway retry loop, executing \texttt{npm install sonner} over 100 times
before intervention.

\textbf{Mitigation:} SPOQ monitors for repeated identical commands (5+
occurrences) and halts with a human consultation request. Organizations
should implement:
\begin{itemize}
    \item Command deduplication with exponential backoff
    \item Per-task execution time limits (default: 30 minutes)
    \item Anomaly detection for unusual command patterns
\end{itemize}

\paragraph{Validation Gaming.}
Agents optimizing for validation metrics may produce code that passes
automated checks while failing to address underlying requirements. Examples
include: tests that assert \texttt{true === true}, implementations that
short-circuit edge cases, and documentation that restates function signatures
without explaining behavior.

\textbf{Mitigation:} The dual validation framework addresses surface-level
gaming through complementary metrics (test existence vs. test pass rate,
syntactic correctness vs. requirements fidelity). However, sophisticated
gaming requires human oversight:
\begin{itemize}
    \item Periodic human code review of high-complexity tasks
    \item Mutation testing to verify test quality
    \item Requirements traceability audits
    \item Cross-validation between independent agents on critical paths
\end{itemize}

\paragraph{Dependency Resolution Failures.}
Agents may specify incorrect dependency versions, introduce circular
dependencies, or fail to recognize version conflicts. These failures often
manifest only at runtime or during integration testing.

\textbf{Mitigation:}
\begin{itemize}
    \item Lock files (\texttt{package-lock.json}, \texttt{Cargo.lock}) committed
    after Wave 0 dependency installation
    \item Dependency audit as part of code validation metrics
    \item Version pinning policies enforced via task templates
    \item Integration tests run at wave boundaries to detect compatibility issues
\end{itemize}

\needspace{4cm}
\subsubsection{Coordination Failures}
\label{subsubsec:coordination-detail}

\paragraph{Agent Coordination Failures.}
In parallel execution, agents may make conflicting assumptions about shared
state. Example scenarios include:
\begin{itemize}
    \item Two agents independently creating the same utility function with
    different signatures
    \item An agent assuming a database table exists while another agent is
    still creating it
    \item Conflicting CSS class names or component identifiers
\end{itemize}

\textbf{Mitigation:} Wave boundaries serve as synchronization points.
Additional measures include:
\begin{itemize}
    \item Explicit interface contracts defined in task prerequisites
    \item Naming conventions specified in task templates
    \item Linting rules to detect common conflict patterns
    \item Post-wave merge verification before proceeding
\end{itemize}

\paragraph{Human Consultation Latency.}
When agents request human consultation (HaaA), execution pauses until
response. If the human is unavailable, this creates a bottleneck that
can stall entire wave progressions.

\textbf{Mitigation:}
\begin{itemize}
    \item Configurable timeout with fallback behavior (skip, abort, or
    proceed with best-effort)
    \item Escalation to secondary human reviewers
    \item Asynchronous consultation queues for non-blocking requests
    \item Clear documentation of expected response times per request priority
\end{itemize}

\subsubsection{Cost and Resource Failures}
\label{subsubsec:cost-resource-detail}

\paragraph{Cost Runaway Risk.}
Without limits, agents retrying failed tasks or entering verbose output
loops can rapidly consume API credits. A single runaway agent in Case Study 1
consumed approximately \$15 in tokens before detection. Extrapolated to
parallel execution with 10+ agents, uncontrolled failures could exceed
\$100/hour.

\textbf{Mitigation:}
\begin{itemize}
    \item Per-task token budgets (recommended: 50K input, 10K output for
    standard tasks)
    \item Per-epic cost caps with automatic suspension
    \item Real-time cost monitoring dashboards with alerting
    \item Graduated rate limiting: warning at 80\%, throttle at 90\%, halt
    at 100\%
\end{itemize}

\needspace{4cm}
\subsubsection{Security and Isolation Failures}
\label{subsubsec:security-isolation-detail}

\paragraph{Execution Isolation.}
Agents execute commands in the host environment without sandboxing.
Malformed or malicious code could damage the system, exfiltrate data,
or consume resources. While LLM agents are unlikely to be intentionally
malicious, hallucinated commands (e.g., \texttt{rm -rf /}) pose real risks.

\textbf{Mitigation:}
\begin{itemize}
    \item Container-based execution (Docker) with ephemeral workspaces
    \item Read-only filesystem mounts where possible
    \item Network isolation for tasks that do not require external access
    \item Resource limits (CPU, memory, disk quotas)
    \item Command allowlisting for high-risk operations
    \item Audit logging of all executed commands
\end{itemize}

\paragraph{Rollback and Recovery.}
If agent work corrupts the codebase, recovery requires manual git operations.
SPOQ does not currently implement automatic rollback, relying instead on
git history and manual intervention.

\textbf{Mitigation:}
\begin{itemize}
    \item Commit checkpoints at wave boundaries, enabling \texttt{git revert}
    to known-good states
    \item Branch-per-epic isolation with squash merge on completion
    \item Automated backup before epic execution
    \item Recovery runbooks documenting rollback procedures for common failures
\end{itemize}

\subsection{Evaluation Tables}
\label{subsec:evaluation-tables}

\needspace{8cm}
\begin{table}[htbp]
\centering
\caption{Operational Risk Summary}
\label{tab:risk-summary}
\begin{tabular}{lll}
\toprule
\textbf{Risk Category} & \textbf{Detection Signal} & \textbf{Mitigation} \\
\midrule
Context exhaustion & Output truncation & Scope limits, context budgets \\
Cost runaway & Token counter spikes & Per-task budgets with alerts \\
Human bottleneck & Queue timeout & Configurable fallback \\
Lock contention & EBUSY errors & Single-agent dep.\ install \\
Validation gaming & High scores, low quality & Human review, mutation testing \\
Agent conflicts & Merge failures & Interface contracts, wave sync \\
Runaway loops & Repeated commands & Deduplication, time limits \\
Security breach & Unauthorized commands & Container isolation, allowlists \\
Data corruption & Test/build failures & Wave checkpoints, rollback \\
\bottomrule
\end{tabular}
\end{table}

\begin{table}[htbp]
\centering
\caption{Evaluation Summary}
\label{tab:summary}
\begin{tabular}{lccc}
\toprule
\textbf{Metric} & \textbf{UI Epic} & \textbf{Rebrand Epic} & \textbf{Adoption (agg.)} \\
\midrule
Tasks & 13 & 12 & 92 \\
Waves & 2 & 4 & varies \\
Max parallelism & 12 & 5 & 4 \\
Speedup factor & 5.3$\times$ & 2.8$\times$ & 1.3--3.0$\times$ \\
Completion rate & 92\% & 100\% & 100\% \\
Orchestrator interventions & 1 & 3 & varies \\
Rework cycles & 2 & 0 & 0--1 \\
Test cases & --- & 174 & 295 \\
Avg. confidence & --- & 0.92 & 0.90--0.95 \\
\bottomrule
\end{tabular}
\end{table}

\clearpage
\subsection{Cost Analysis and ROI Framework}
\label{subsec:cost-analysis-detail}

A critical consideration for SPOQ adoption is the economic viability of
multi-agent orchestration. We analyze costs under two pricing models
available for Claude API access as of February 2025.

\paragraph{Pricing Model A: Per-Token API.}
Table~\ref{tab:api-pricing} presents the current Claude API pricing structure
and how each tier maps to SPOQ agent roles.

\begin{table}[htbp]
\centering
\caption{Claude API Pricing by Model Tier}
\label{tab:api-pricing}
\begin{tabular}{lccl}
\toprule
\textbf{Model} & \textbf{Input} & \textbf{Output} & \textbf{SPOQ Role} \\
\midrule
Opus 4.6 & \$15/M tokens & \$75/M tokens & Worker agents \\
Sonnet 4.5 & \$3/M tokens & \$15/M tokens & Reviewer agents \\
Haiku 4.5 & \$0.25/M tokens & \$1.25/M tokens & Investigator agents \\
\bottomrule
\end{tabular}
\end{table}

Based on observed token consumption in our case studies, a typical Opus worker
task consumes approximately 25,000 input tokens and 5,000 output tokens,
yielding a per-task cost of approximately \$1.95. For an epic of 13 tasks
(similar to our UI improvements study), the total worker cost reaches
approximately \$28, excluding reviewer and investigator overhead.

\paragraph{Pricing Model B: Flat-Rate Max Plan.}
Anthropic's Max plan (\$200/month) provides 20$\times$ the usage allowance
of a standard Claude Pro subscription. There is no platform-imposed limit on
concurrent Claude Code instances; the practical ceiling is human attention.
Crucially, usage is metered in two separate buckets: \emph{Opus} consumption
and \emph{non-Opus} consumption (Sonnet, Haiku). This two-bucket structure
makes SPOQ's three-tier agent hierarchy an economic optimization as well as
a capability one; reserving Opus tokens for task execution while routing
validation and triage through Sonnet and Haiku preserves Opus headroom for
the work that demands it most.

\begin{itemize}
    \item \textbf{Fixed monthly cost} with predictable budgeting
    \item \textbf{Concurrency:} unlimited instances; bounded only by the
    two usage buckets and human supervisory bandwidth
    \item \textbf{Tiered metering:} Opus budget for Workers; Sonnet/Haiku
    budget for Reviewers and Investigators
    \item \textbf{Daily capacity:} 50--100 tasks (assuming 4--6 task cycles
    across active instances)
    \item \textbf{Effective per-task cost:} \$0.10 at scale (100 tasks/day
    $\times$ 20 working days)
\end{itemize}

At scale, the Max plan reduces per-task costs by approximately 20$\times$
compared to per-token pricing, making aggressive parallelization economically
viable on a single subscription.

\clearpage
\paragraph{The Director Model.}
We propose a scaling paradigm we term the \emph{Director Model}, wherein a
single human engineer orchestrates multiple Claude Code instances under one
Max license:

\begin{itemize}
    \item \textbf{1 planning instance:} Assists with epic decomposition and
    dependency analysis
    \item \textbf{5 execution instances:} Each runs SPOQ's wave-based dispatch
    with parallel sub-agents
    \item \textbf{Daily output:} 50--100 completed tasks
\end{itemize}

In our experience, six concurrent instances represents the human
multitasking limit; the usage buckets are rarely exhausted even at this
level, leaving headroom for sustained parallel execution. This configuration
enables a single engineer
to achieve throughput equivalent to 5--8 traditional engineers, representing
a qualitative shift in individual productivity potential.

\paragraph{ROI Framework.}
Table~\ref{tab:roi} presents a monthly ROI calculation for the Director Model
at typical utilization.

\begin{table}[htbp]
\centering
\caption{Director Model Monthly ROI Estimate}
\label{tab:roi}
\begin{tabular}{lr}
\toprule
\textbf{Component} & \textbf{Value} \\
\midrule
Monthly tasks completed & 2,000 (100/day $\times$ 20 days) \\
Equivalent engineer output & 5--8 engineers \\
Traditional cost (8 engineers @ \$12,500/mo) & \$100,000 \\
Director Model cost (1 Max license) & \$200 \\
Engineer salary (director) & \$12,500 \\
Total Director Model cost & \$12,700 \\
\midrule
\textbf{Monthly savings} & \$87,300 \\
\textbf{ROI multiplier} & 7.9$\times$ \\
\bottomrule
\end{tabular}
\end{table}

Under favorable assumptions (consistent utilization, stable task completion),
the Director Model yields approximately 8$\times$ cost efficiency compared
to traditional staffing.

\paragraph{Caveats and Variability.}
These estimates carry significant uncertainty:

\begin{enumerate}
    \item \textbf{Task complexity variance:} Our cost-per-task figures derive
    from 1--4 hour tasks. Complex tasks requiring extended context windows or
    multiple rework cycles can cost 3--5$\times$ the baseline.

    \item \textbf{Rework overhead:} Failed validations requiring remediation
    add approximately 40\% overhead on affected tasks.

    \item \textbf{Output quality equivalence:} The ROI comparison assumes
    agent-completed tasks are comparable in quality to human-authored work.
    This equivalence has not been empirically validated; agent output may
    require additional review or refinement that narrows the cost gap.

    \item \textbf{Human supervision cost:} The Director Model assumes skilled
    engineers capable of effective agent orchestration. Training and context-switching
    costs are not included.

    \item \textbf{API rate limits:} Per-token pricing may encounter rate limits
    at high parallelism, while Max plan throughput is bounded by
    the usage bucket allowances under sustained heavy parallelism.

    \item \textbf{Quality vs. speed tradeoffs:} Higher parallelism may introduce
    integration issues that offset time savings with debugging overhead.
\end{enumerate}

Despite these caveats, the order-of-magnitude cost advantage suggests that
multi-agent orchestration represents an economically viable paradigm for
software development at scale.

\needspace{6cm}
\subsection{When to Use SPOQ: Decision Framework}
\label{subsec:usage-guidance-detail}

SPOQ's structured approach introduces upfront overhead (task
decomposition, YAML specification, dependency mapping, and validation
gates) that pays off for certain project types but hinders others.
This subsection provides decision guidance for practitioners.

\paragraph{Appropriate Use Cases.}
SPOQ adds value when:

\begin{itemize}
    \item \textbf{Structured features:} New functionality with clear
    requirements spanning multiple components. When you can articulate
    what needs to be built before starting, SPOQ's planning investment
    translates into parallel execution benefits.

    \item \textbf{Multi-file refactors:} Coordinated changes across a
    codebase where parallelism accelerates delivery. Renaming a widely-used
    interface, migrating to a new library, or restructuring module
    boundaries benefit from explicit dependency tracking.

    \item \textbf{Infrastructure work:} Terraform configurations, CI/CD
    pipelines, and deployment automation have well-defined outputs
    and benefit from explicit dependency specification.

    \item \textbf{Team handoffs:} Projects requiring audit trails,
    explainability for stakeholders, or handoffs between engineers
    benefit from SPOQ's journal tracking and task documentation.

    \item \textbf{Scale thresholds:} Our empirical data suggests SPOQ
    provides net benefit for epics with 5+ tasks spanning 4+ hours
    of estimated work. Below this threshold, orchestration overhead
    often exceeds execution time.
\end{itemize}

\paragraph{Inappropriate Use Cases.}
SPOQ's overhead exceeds its value when:

\begin{itemize}
    \item \textbf{Exploratory prototyping:} When requirements are
    unclear and the goal is discovery, SPOQ's upfront planning becomes
    a burden. Rapid iteration with a single agent or direct coding
    serves exploration better.

    \item \textbf{Small fixes:} Single-file bug fixes or minor changes
    where task decomposition time exceeds execution time. A 10-minute
    fix should not require 30 minutes of specification.

    \item \textbf{Creative flow states:} Sessions where the developer
    seeks direct engagement with code, what practitioners sometimes
    call ``vibe coding'', benefit from immediacy, not delegation.
    SPOQ assumes you want to orchestrate, not implement.

    \item \textbf{Urgent hotfixes:} Time-critical patches where
    validation gates delay resolution. When production is down,
    skip the methodology and fix the problem directly.

    \item \textbf{Learning exercises:} When the goal is understanding
    a codebase or technology, doing the work yourself provides
    education that delegation cannot.
\end{itemize}

\paragraph{Decision Framework.}
A simple heuristic: if specifying tasks and dependencies would take
longer than executing the work directly, SPOQ is likely overkill.
Conversely, if you can identify 5+ subtasks with clear deliverables
and the total estimated effort exceeds 4 hours, SPOQ's parallelism
and quality gates typically justify the planning investment.

Table~\ref{tab:when-to-use} summarizes common scenarios:

\begin{table}[htbp]
\centering
\caption{SPOQ Applicability by Project Type}
\label{tab:when-to-use}
\footnotesize
\begin{tabular}{lcp{6.5cm}}
\toprule
\textbf{Project Type} & \textbf{Use SPOQ?} & \textbf{Rationale} \\
\midrule
New feature (5+ files) & Yes & Parallelism benefits; quality gates prevent rework \\
Single-file bug fix & No & Overhead exceeds benefit \\
Exploratory prototype & No & Requirements unclear; planning premature \\
Migration/refactor & Yes & Coordination critical; dependency tracking valuable \\
Learning exercise & No & Flow state preferred; education through doing \\
Production hotfix & No & Urgency trumps process \\
Infrastructure epic & Yes & Well-defined outputs; explicit dependencies \\
Documentation update & Depends & Multi-file: yes; single page: no \\
\bottomrule
\end{tabular}
\end{table}

\paragraph{Overhead vs.\ Payoff.}
SPOQ's overhead includes: (1) epic planning and task decomposition
(30--90 minutes), (2) YAML specification (5--10 minutes per task),
(3) epic validation iteration (10--30 minutes), and (4) journal
tracking and review (ongoing). For a 10-task epic, expect 2--3 hours
of orchestration overhead.

The payoff comes from: (1) parallel execution reducing wall-clock time,
(2) validation gates catching issues before they cascade, (3) reduced
context-switching as agents handle implementation details, and
(4) audit trails simplifying review and handoffs. Across our case studies
and adoption survey, speedups ranged from 1.3$\times$ to 5.3$\times$
depending on dependency structure and parallelization potential.

The break-even point varies by developer productivity and agent
capability, but our experience suggests epics under 5 tasks rarely
justify SPOQ's overhead, while epics over 15 tasks almost always do.
The 5--15 task range requires judgment about parallelization potential
and team coordination needs.

\end{document}